\documentclass[%
  reprint,
superscriptaddress,
showpacs,preprintnumbers,
nofootinbib,
 amsmath,amssymb, 
 aps,
 prd,
 longbibliography,
]{revtex4-1}

\usepackage{cancel}
\usepackage{accents}
\usepackage{mciteplus,slashed}
\usepackage{amssymb,cancel,amsmath}
\usepackage{dcolumn}
\usepackage{bm}
\usepackage[caption=false]{subfig}
\usepackage{appendix}
\usepackage{feynmp-auto}
\usepackage{tikz-feynman}
\usepackage[T1]{fontenc}	
\usepackage{csvsimple}
\usepackage[colorlinks=true,citecolor=blue,linkcolor=blue, allcolors=blue]{hyperref}
\usepackage[section]{placeins}
\usepackage[capitalise]{cleveref}
\usepackage{booktabs}
\usepackage{graphicx}
\usepackage{mathrsfs}
\usepackage[utf8]{inputenc}
\usepackage{siunitx}
\usepackage{blindtext}

\DeclareSIUnit\electronvolt{e\kern-.05em V}

\begin{document}

\preprint{\hfill FERMILAB-PUB-21-189-AE-T}
\preprint{\hfill MIT-CTP/5298}

\title{New Thermal Relic Targets for Inelastic Vector-Portal Dark Matter}

\author{Patrick J. Fitzpatrick}
\email{fitzppat@mit.edu}
\affiliation{Center for Theoretical Physics, Massachusetts Institute of Technology, Cambridge, Massachusetts 02139, U.S.A.}

\author{Hongwan Liu}
\email{hongwanl@princeton.edu}
\affiliation{Center for Cosmology and Particle Physics, Department of Physics, New York University, New York, NY 10003, U.S.A.}
\affiliation{Department of Physics, Princeton University, Princeton, New Jersey, 08544, U.S.A.}

\author{Tracy R. Slatyer}
\email{tslatyer@mit.edu}
\affiliation{Center for Theoretical Physics, Massachusetts Institute of Technology, Cambridge, Massachusetts 02139, U.S.A.}

\author{Yu-Dai Tsai}
\email{ytsai@fnal.gov}
\affiliation{Fermilab, Fermi National Accelerator Laboratory, Batavia, IL 60510, U.S.A.}

\date{\today}

\begin{abstract}


We examine the vector-portal inelastic dark matter (DM) model with DM mass $m_\chi$ and dark photon mass $m_{A'}$, in the `forbidden dark matter' regime where $1 \lesssim m_{A'}/m_\chi \lesssim 2$, carefully tracking the dark sector temperature throughout freezeout. The inelastic nature of the dark sector relaxes the stringent cosmic microwave background (CMB) and self-interaction constraints compared to symmetric DM models. We determine the CMB limits on both annihilations involving excited states and annihilation into $e^+e^-$ through initial-state-radiation of an $A'$, as well as limits on the DM self-scattering, which proceeds at the one-loop level. The unconstrained parameter space serves as an ideal target for accelerator $A'$ searches, and provides a DM self-interaction cross section that is large enough to observably impact small-scale structure.


\end{abstract}
 
\maketitle 

\textbf{Introduction.---} Light dark matter (DM) composed of particles in the \SI{}{\mega\eV}--\SI{}{\giga\eV} range has received considerable attention in recent years, with significant progress made in terms of model building, its production in the early Universe, and detection strategies (see Ref.~\cite{Battaglieri:2017aum} for a recent community report summarizing some of these developments). One of the most common benchmark models is the vector or kinetic mixing portal, in which the DM particle $\chi$ interacts with the Standard Model (SM) through a dark photon mediator $A'$, which in turn kinetically mixes with the SM photon~\cite{Holdom:1985ag}. For sufficiently large kinetic mixing parameter $\epsilon$, the dark sector thermalizes with the SM while dark sector particles are still relativistic. In this scenario, the ratio of the dark photon mass to the DM mass, $r \equiv m_{A'}/m_\chi$, divides the production mechanisms of DM into several distinct regimes. For $r > 2$, DM undergoes a thermal freezeout via annihilation into a pair of SM particles, while for $r < 1$, DM freezes out via a secluded annihilation~\cite{Pospelov:2007mp} into two $A'$s. In either case, if the annihilation rate of DM is unsuppressed during recombination relative to the annihilation rate during freezeout, the Planck cosmic microwave background (CMB) power spectrum rules out the large annihilation rates required to produce the correct DM relic abundance~\cite{Aghanim:2018eyx}. This limit is not fully model-independent; for example, it can be relaxed if the DM is asymmetric. For $r > 2$, the annihilation of scalar or Majorana DM is dominantly $p$-wave and so is suppressed at low velocities, naturally evading the constraints (see, e.g., Ref.~\cite{Battaglieri:2017aum}).

The regime with $1 \lesssim r \lesssim 2$ exhibits a rich freezeout phenomenology, allowing for a symmetric DM population with an s-wave annihilation cross section. Under the assumption that the dark sector remains thermally coupled to the SM throughout freezeout, Refs.~\cite{PhysRevLett.115.061301, Cline:2017tka} studied the freezeout of symmetric, Dirac fermionic DM $\chi$ for $1 \lesssim r \lesssim 2$, finding unconstrained parameter space for this model that provides interesting search targets for beam and direct detection experiments. This model also generically requires a substantial dark sector coupling between $\chi$ and $A'$ for the correct relic abundance to be achieved, giving DM a significant self-interaction rate, which can modify dark matter structures on small scales.

In Ref.~\cite{Fitzpatrick:2020vba}, we revisited the $1 \lesssim r \lesssim 2$ regime, this time lifting the assumption of efficient thermal contact between the dark sector and the SM. We showed that dark sector processes like $\chi \overline{\chi} \leftrightarrow A'A'$ (hereafter denoted $2 \leftrightarrow 2$) and $\chi \overline{\chi} \chi \leftrightarrow \chi A'$ (hereafter denoted $3 \leftrightarrow 2$) can generate or remove a significant amount of heat during freezeout; if the energy transfer rate between the SM and the dark sector is not large enough, this can lead to the dark sector kinetically decoupling from the SM before freezeout is complete, leading to a dark sector temperature $T'$ that is different from the SM temperature $T$. Kinetic decoupling can occur before freezeout is complete for $\epsilon$ as large as $10^{-5}$ for $r \lesssim 1.5$, dramatically altering how the correct relic abundance is achieved. While some parameter space in the range $\SI{100}{\mega\eV} \lesssim m_\chi \lesssim \SI{1}{\giga\eV}$ remains open, especially for $r \lesssim 1.5$, this model still faces strong constraints from limits on the self-interaction cross section of DM, as well as the aforementioned CMB power spectrum. 

In this Letter, we investigate the experimental constraints on \textit{inelastic} DM in the $1 \lesssim r \lesssim 2$ regime, where the dark matter is now made up of a Majorana ground state $\chi$ and excited state $\chi^*$, with a small mass splitting between them~\cite{TuckerSmith:2001hy}, with coupling to the $A'$ now occurring off-diagonally, i.e., only between the $\chi$ and $\chi^*$ states. The existence of the mass splitting suppresses the primordial abundance of $\chi^*$ relative to $\chi$, decreasing the annihilation rate of $\chi$ during recombination and lifting the CMB power spectrum constraints. The off-diagonal nature of the coupling ensures that self-interaction between $\chi$ particles is forbidden at tree-level, reducing the severity of the self-interaction constraints. This model has a significantly enlarged range of experimentally allowed parameter space, including a new window at $m_\chi \sim \SI{10}{\mega\eV}$ and $\epsilon \sim 10^{-8}$. Our results motivate future beam experiments to close the full range of $\epsilon$ over which the dark sector can thermalize with the SM in the early Universe, as well as an improved understanding of supernova cooling constraints on light DM at small mixing. 

In the remainder of this Letter, we will review the inelastic DM model, the predicted primordial abundance of the excited state, and the newly relevant CMB and self-interaction constraints that replace those applicable to a symmetric DM model. We will then conclude by examining the existing experimental constraints on this model. More details of the inelastic DM model, a discussion of subdominant or model-dependent constraints, and the details of our calculation of the $\chi \chi \to \chi \chi$ one-loop self-interaction cross section can be found in the Supplemental Material. 

\textbf{Vector-portal inelastic dark matter.---} Our dark sector contains a dark photon $A'$ with mass $m_{A'}$, the massive gauge boson of a broken U(1) gauge symmetry in the dark sector, as well as a pair of Majorana fermions $\chi$ and $\chi^*$ that makes up the DM, which we will refer to as the `ground state' and `excited state' respectively. Both states have similar masses $m_\chi$ and $m_{\chi^*}$, with a small dimensionless mass splitting $\delta$ defined by $m_{\chi^*} - m_\chi \equiv \delta m_\chi$, and $\delta \ll 1$.

Due to the ultraviolet (UV) origins of the mass splitting and the symmetry breaking of the dark U(1), which we will describe in more detail below, the dark fermions couple off-diagonally to $A'$. Moreover, $A'$ kinetically mixes with the Standard Model (SM) photon~\cite{Holdom:1985ag}, generating a coupling between the SM electromagnetic current $J^\mu_\text{EM}$ and $A'$. The terms in the Lagrangian chiefly responsible for the phenomenology of this model are thus
\begin{alignat}{1}
    \mathcal{L} \supset  g_D A_\mu' \overline{\chi^*} \gamma^\mu \chi + \epsilon e A_\mu' J^\mu_\text{EM} \,,
\end{alignat}
where $\epsilon$ is the kinetic mixing parameter, $g_D$ is the dark sector coupling and $e$ is the electron charge. 

The small mass splitting $\delta$, the dark fermion masses and the breaking of the U(1) symmetry associated with $A'$ can all be simultaneously achieved by a judicious choice of the symmetry breaking pattern and couplings between a dark Higgs field $\Phi$ and a Dirac fermion $\Psi$, which is split into $\chi$ and $\chi^*$ mass eigenstates after symmetry breaking. Many models to achieve this have been proposed; two specific example models~\cite{Finkbeiner:2010sm,Elor:2018xku} are discussed in the Supplemental Material. In the limit where the dark Higgs particle $h_D$ is much more massive than $\chi$, there is no significant difference in the phenomenology of the dark sector between these two models; by default we will assume this condition.

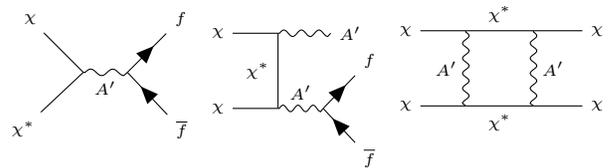
\begin{figure}[t!]
    \centering
    \raisebox{-0.55\height}{\begin{tikzpicture}
        \begin{feynman}
            \vertex(i1) {\(\scriptstyle{\chi}\)}; 
            \vertex[below right=1cm of i1] (a);
            \vertex[below left=0.75cm of a]  (i2) {\(\scriptstyle{\chi^*}\)};

            \vertex[right=2cm of i1] (f1) {\(\scriptstyle{f}\)};
            \vertex[below left=1cm of f1]  (c);
            \vertex[below right=0.75cm of c]  (f2) {\(\scriptstyle{\overline{f}}\)};

            \diagram*{
                (i1) -- (a) -- (i2),
                (a) -- [boson, edge label'=\(\scriptstyle{A'}\)] (c),
                (f2) -- [fermion] (c) -- [fermion] (f1)
            };
        \end{feynman}
    \end{tikzpicture}}
    \raisebox{-0.68\height}{\begin{tikzpicture}
        \begin{feynman}
            \vertex (i1) {\(\scriptstyle{\chi}\)}; 
            \vertex[right=0.8cm of i1] (a);
            \vertex[right=0.7cm of a]  (f1) {\(\scriptstyle{A'}\)};

            \vertex[below=1cm of i1] (i2) {\(\scriptstyle{\chi}\)};
            \vertex[right=0.8cm of i2] (c);
            \vertex[right=0.6cm of c] (d);
            \vertex[above right=0.6cm of d]  (f2) {\(\scriptstyle{f}\)};
            \vertex[below right=0.6cm of d]  (f3) {\(\scriptstyle{\overline{f}}\)};

            \diagram*{
                (i1) -- (a) -- [boson] (f1),
                (a) -- [edge label'=\(\scriptstyle{\chi^*}\)] (c),
                (i2) -- (c),
                (c) -- [boson, edge label=\(\scriptstyle{A'}\)] (d),
                (d) -- [fermion] (f2),
                (f3) -- [fermion] (d),
            };
        \end{feynman}
    \end{tikzpicture}}
    \raisebox{-0.5\height}{\begin{tikzpicture}
        \begin{feynman}
            \vertex(i1) {\(\scriptstyle{\chi}\)}; 
            \vertex[right=0.8cm of i1] (a);
            \vertex[right=0.9cm of a]  (b);
            \vertex[right=0.65cm of b]  (f1) {\(\scriptstyle{\chi}\)};

            \vertex[below=1cm of i1] (i2) {\(\scriptstyle{\chi}\)};
            \vertex[right=0.8cm of i2] (c);
            \vertex[right=0.9cm of c]  (d);
            \vertex[right=0.65cm of d]  (f2) {\(\scriptstyle{\chi}\)};

            \diagram*{
                (i1) -- (a) -- [edge label=\(\scriptstyle{\chi^*}\)] (b),
                (b) -- (f1),
                (i2) -- (c) -- [edge label'=\(\scriptstyle{\chi^*}\)] (d),
                (d) -- (f2),
                (a) -- [boson, edge label'=\(\scriptstyle{A'}\)] (c),
                (b) -- [boson, edge label=\(\scriptstyle{A'}\)] (d),
            };
        \end{feynman}
    \end{tikzpicture}}
    \caption{Feynman diagrams for important processes in the inelastic vector-portal dark matter model with $1 \lesssim r \lesssim 2$: (left) annihilation of DM with primordially produced excited states, $\chi^* \chi \to f \overline{f}$, where $f$ is a SM fermion; (center) annihilation of DM with initial state radiation, $\chi \chi \to A' A'^*$, $A'^* \to f \overline{f}$, and (right) DM self-interaction, $\chi \chi \to \chi \chi$ (only one diagram shown here: see Supplemental Material for a complete discussion).}
    \label{fig:feynman_diagrams}
\end{figure}

There are several tree-level processes present in the symmetric model considered in Ref.~\cite{Fitzpatrick:2020vba} that are absent in this inelastic case: DM ground-state annihilation into SM particles, tree-level scattering between the DM and SM fermions, and DM self-interactions. This relaxes the resulting constraints significantly. In Fig.~\ref{fig:feynman_diagrams}, we show three processes that are of particular importance to the inelastic vector-portal DM model, which we will now consider in turn.

\textbf{Freezeout and primordial excited state abundance.---} For $\epsilon \gtrsim 10^{-9}$, the dark sector achieves thermal equilibrium with the SM while dark sector particles are relativistic~\cite{Fitzpatrick:2020vba}. Furthermore, while $T'/m_\chi \gg \delta$ (where $T'$ is the dark sector temperature), the mass splitting between $\chi$ and $\chi^*$ is irrelevant, and the dark fermions can equivalently be treated as part of the Dirac fermion $\Psi$. Taking $\delta \lesssim 10^{-3}$ ensures that this condition holds throughout freezeout. In this case, the results of Ref.~\cite{Fitzpatrick:2020vba} are fully applicable to the evolution of the dark sector until all DM number-changing processes have frozen out, fixing the DM abundance. After this point, the only remaining processes that are fast compared to cosmic expansion involve total-number-conserving conversions between $\chi$ and $\chi^*$ only.

After the freezeout of DM, $\chi$ and $\chi^*$ particles stay in chemical equilibrium as the dark sector cools, until $T'/m_\chi \sim \delta$, when the number density of $\chi^*$ starts becoming Boltzmann-suppressed relative to the number density of $\chi$. Eventually, the comoving number density of $\chi^*$ freezes out when the number-changing process $\chi^*\chi^* \leftrightarrow \chi \chi$ becomes slow, i.e., when $n_{\chi^*} \langle \sigma v \rangle_{\chi^* \chi^* \to \chi \chi} \sim H$. The annihilation cross section is given by~\cite{Baryakhtar:2020rwy}
\begin{alignat}{1}
    \langle \sigma v \rangle_{\chi^* \chi^* \to \chi \chi} \simeq \frac{8 \sqrt{\pi}}{r^4} \frac{\alpha_D^2}{m_\chi^2} \times 
    \begin{cases}
        \sqrt{1/x'} \,, & \delta \ll 1/x' \,, \\
        \sqrt{\pi \delta/2} \,, & \delta \gg 1/x' \,,
    \end{cases}
\end{alignat}
where $x' \equiv m_\chi/T'$, and $\alpha_D \equiv g_D^2 / (4\pi)$. In order for $\chi$ to make up all of the DM, $m_\chi n_\chi \sim T_\text{eq} T^3$, where $T_\text{eq}$ is the temperature of the CMB at matter-radiation equality. Thus we can estimate the ratio $n_{\chi^*}/n_\chi$ to be
\begin{alignat}{1}
    \frac{n_{\chi^*}}{n_\chi} \sim \frac{1}{4 \sqrt{2} \pi \sqrt{\delta}} \frac{r^4 x_*}{M_\text{pl} T_\text{eq}} \frac{m_\chi^2}{\alpha_D^2} \,,
\end{alignat}
where $x_*$ ($x_*'$) is defined by the SM (dark sector) temperature at which $\chi^* \chi^* \leftrightarrow \chi \chi$ freezes out. Since the dark sector temperature redshifts as $(1+z)^2$ after DM freezes out, which occurs at SM and dark sector temperatures $x_f$ and $x_f'$ respectively, we can write $x_* \approx x_f (x_*'/ x_f')^{1/2}$, giving this parametric expression for $n_{\chi^*}/n_\chi$: 
\begin{multline}
    \frac{n_{\chi^*}}{n_\chi} \sim 10^{-7} \left(\frac{m_\chi}{\SI{10}{\mega \eV}}\right)^2 \left(\frac{r}{1.6}\right)^4 \left(\frac{0.1}{\alpha_D}\right)^2 \\
    \times \left(\frac{x_f}{200}\right) \left(\frac{10}{x_f'}\right)^{1/2} \left(\frac{x_*' \delta}{20}\right)^{1/2} \left(\frac{10^{-3}}{\delta}\right) \,,
\end{multline}
keeping in mind that $x_*' \delta$ is the ratio of the mass splitting to the DM temperature at $\chi^* \chi^* \leftrightarrow \chi \chi$ freezeout. We see that $n_{\chi^*}$ can be easily suppressed by seven orders of magnitude relative to $n_\chi$ in our parameter space of interest.

Even with this small primordial abundance, $\chi^* \chi$ annihilations into energetic SM particles are potentially constrained by the CMB power spectrum~\cite{Aghanim:2018eyx}. The limit is given approximately as $f(m_\chi) \langle \sigma v \rangle_{\chi^* \chi \to e^+e^-} \lesssim \SI{1.7e-30}{\centi\meter\cubed\per\second} (m_\chi / \SI{10}{\mega\eV}) n_\chi/n_{\chi^*} $, where $f(m_\chi)$ is an efficiency factor accounting for delayed absorption of energy injected through annihilations~\cite{Aghanim:2018eyx,Slatyer:2015jla}.

The decays of $\chi^*$ through an off-shell $A'$ to the ground state can produce high-energy particles that may be constrained by their  effect on the CMB power spectrum and primordial elemental abundances from Big Bang Nucleosynthesis (BBN)~\cite{Poulin:2016anj}. For $\delta \lesssim 2 \times 10^{-4}$, these constraints are unimportant in the parameter space of interest, and so for convenience we choose $\delta = 10^{-4}$ in this Letter. For a discussion of these decays and the relevant constraints at larger values of $\delta$, we refer the reader to the Supplemental Material.

\textbf{CMB initial state radiation limits.---} Even without an excited state population during recombination, DM self-annihilation with initial state radiation (ISR) $\chi \chi \to A' A'^*$, $A'^* \to f \overline{f}$ is kinematically allowed, with one $A'$ produced off-shell. This process is particularly important in the forbidden regime~\cite{Rizzo:2020jsm}, and may also be constrained by the CMB power spectrum (we also considered $\chi \chi \to f \overline{f}$ annihilations at one-loop, but find they are subdominant to both ISR and the $\chi \chi^*$ annihilations discussed above). In our model, the annihilation cross section into $e^+e^-$ is given by~\cite{Rizzo:2020jsm}
\begin{multline}
    \langle \sigma v \rangle_{\chi \chi \to A'e^+e^-} = \SI{1.3e-35}{\centi\meter\cubed\per\second} \left(\frac{1.6}{r}\right)^2\\
    \times \left(\frac{\alpha_D}{0.1}\right)^2 \left(\frac{\epsilon}{10^{-8}}\right)^2 \left(\frac{\SI{10}{\mega\eV}}{m_\chi}\right)^2 \left(\frac{I(r,\delta)}{10^{-2}}\right) \,,
\end{multline}
where $I(r,\delta)$ is a phase space integral that ranges from approximately $10^{-3}$ to $0.25$ in the range $1.2 \lesssim r \lesssim 1.8$; the fit $\log_{10} I(r,\delta) \approx -0.19 + 5.0 \log_{10}(2-r) + 1.7 \log_{10}^2(2-r)$ is accurate to within 20\% in the same range of $r$-values, as long as $\delta \lesssim 0.1$. Comparing this expression with the CMB limit of approximately $ f(m_\chi) \langle \sigma v \rangle \lesssim \SI{3.3e-30}{\centi\meter\cubed\per\second} (m_\chi / \SI{10}{\mega\eV})$, we find that the CMB power spectrum limits on energy injection offer only mild constraints in our parameter space of interest. 

\textbf{Self-interaction limits.---} The other key constraint on this model comes from self-interaction between $\chi$ particles, which can modify the structure of galaxies; we adopt an upper limit of $\sigma/m_\chi < \SI{1}{\centi\meter\squared\per\gram}$ (e.g., Ref.~\cite{Bondarenko:2020mpf}) on this cross section. $\chi \chi \rightarrow \chi \chi$ elastic scattering via $A'$ exchange is forbidden at tree-level, but there is a non-zero contribution at one-loop order. We compute this one-loop cross section in the low-velocity limit; in the range $1 \lesssim r \lesssim 2$, the cross section is well-approximated numerically by 
\begin{alignat}{1}
    \sigma_{\chi\chi} \approx \frac{3}{4} \left(\frac{\pi}{2 r^6} + \frac{18}{\pi r^4}\right) \frac{\alpha_D^4}{m_\chi^2} \,.
    \label{eq:1loop}
\end{alignat}
The complete expression for all $r$ can be found in the Supplementary Materials; this expression agrees with the low-velocity cross section in the massless $A'$ limit derived in Ref.~\cite{Schutz:2014nka}. 

Scattering processes involving initial $\chi^*$ particles can occur (e.g., $\chi \chi^* \rightarrow \chi \chi^*$, $\chi^* \chi^* \rightarrow \chi \chi$) but their rates are suppressed in galaxies by the small abundance of $\chi^*$ in the late Universe. Inelastic scattering can in principle occur from the ground state, $\chi \chi \rightarrow \chi^* \chi^*$, but will be kinematically forbidden provided that $\sqrt{2 \delta}$ exceeds the local maximum velocity (e.g., for $\delta \gtrsim 10^{-4}$, $\sqrt{2\delta}$ is above the typical escape velocity of galaxy clusters). For these reasons, we take the overall self-interaction rate to be given by Eq.~\eqref{eq:1loop}. Compared to the tree-level self-interaction cross section in the symmetric case~\cite{Fitzpatrick:2020vba}, there is a parametric suppression of order $\alpha_D^2$,
which relaxes the constraints on the $m_\chi$-$\epsilon$ parameter space significantly.

There can additionally be a model-dependent tree-level $\chi\chi$ self-interaction via the dark Higgs; the rate will be suppressed by $1/m_{h_D}^4$, where $m_{h_D}$ is the dark Higgs mass, in addition to any model-dependent factors (e.g., one of our example models gives a $\delta^4$ suppression). While this self-interaction rate can in principle dominate for specific models and a sufficiently light dark Higgs, we can safely choose model parameters for this to be subdominant to the one-loop expression in Eq.~\eqref{eq:1loop} without affecting the rest of the analysis. We leave a detailed discussion of the dark Higgs self-interaction rate to the Supplemental Material. 

\begin{figure*}
    \centering
    \includegraphics[width=0.47\textwidth]{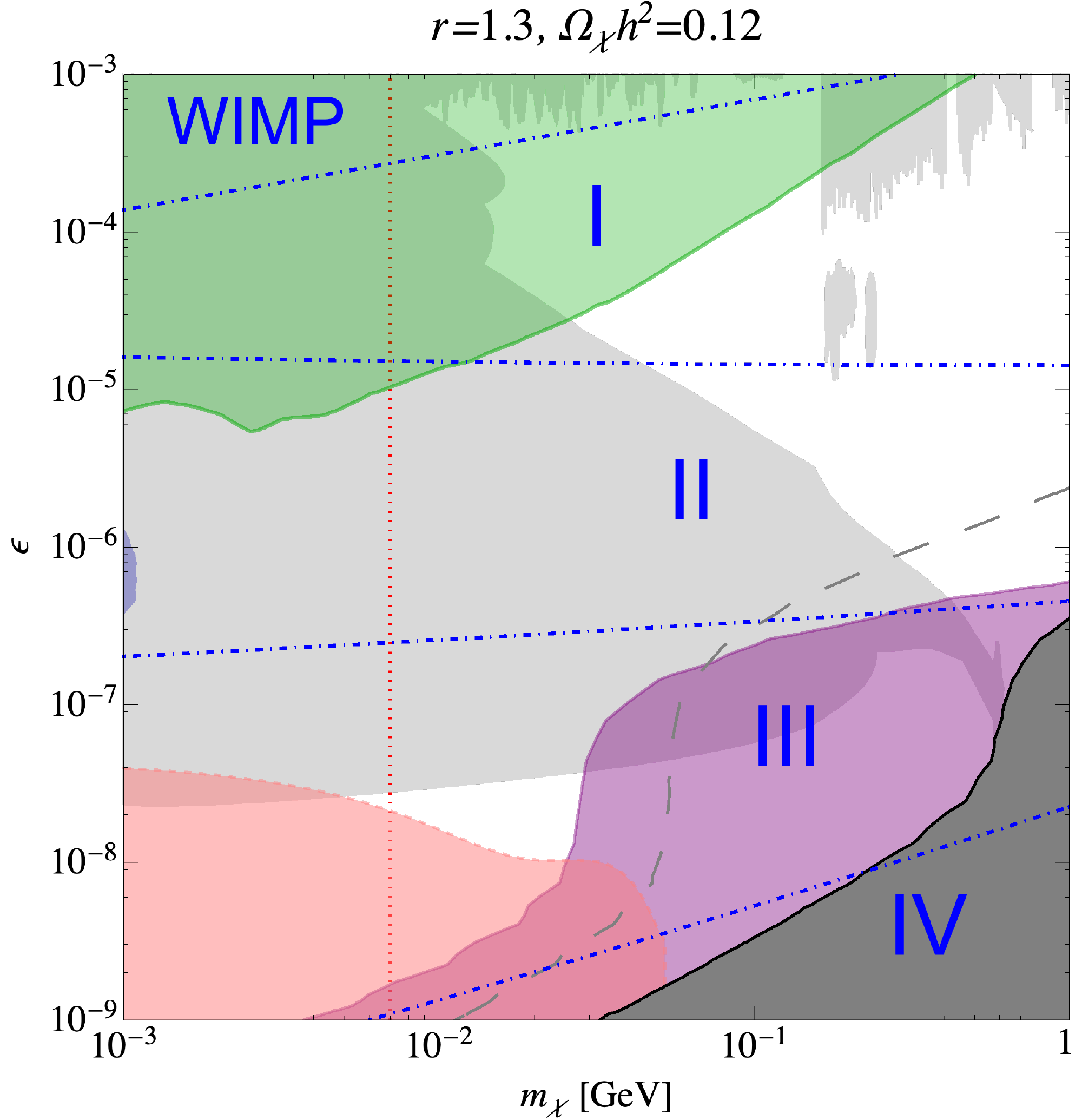}
    \includegraphics[width=0.47\textwidth]{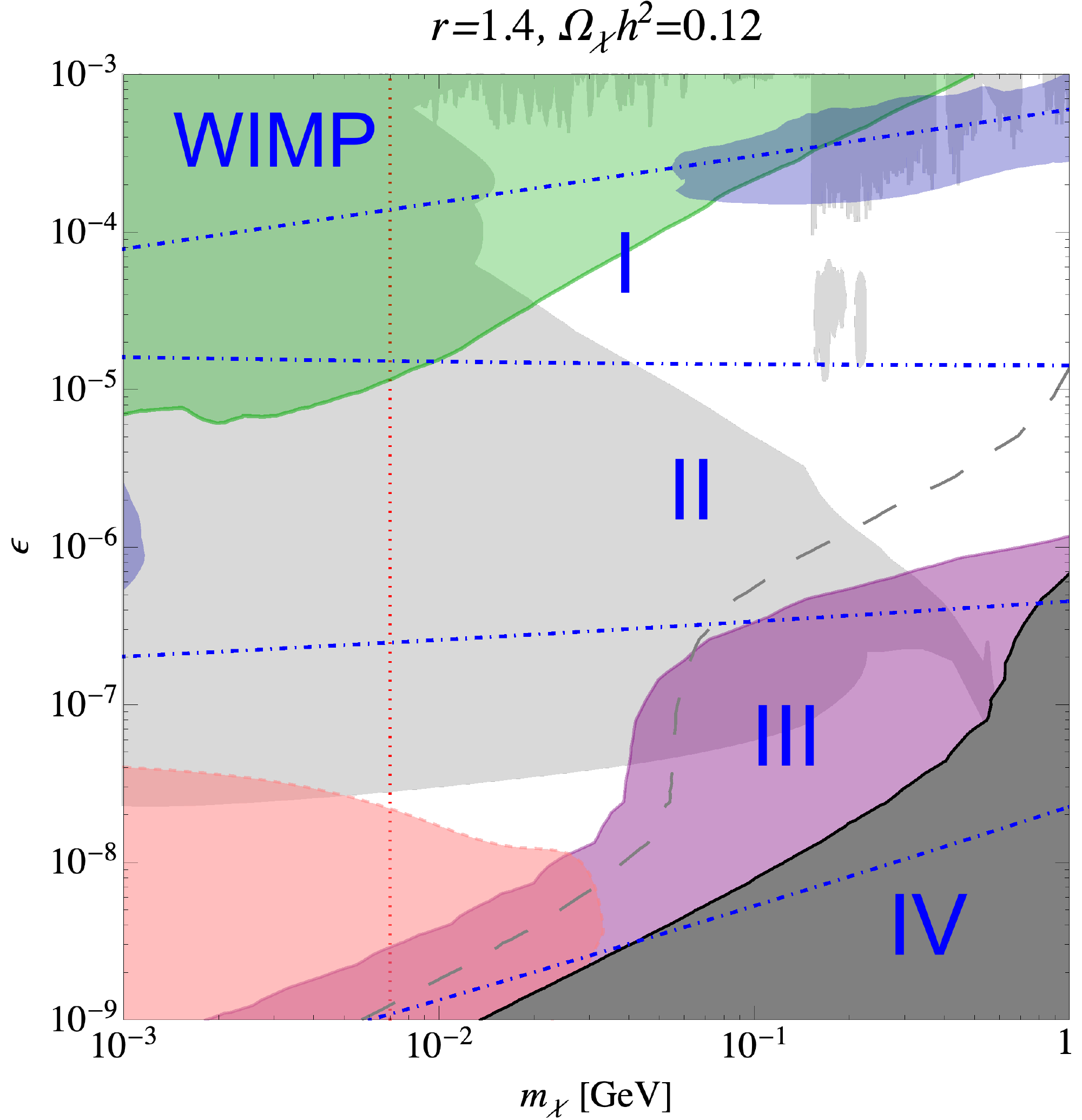}
    \qquad
    \includegraphics[width=0.47\textwidth]{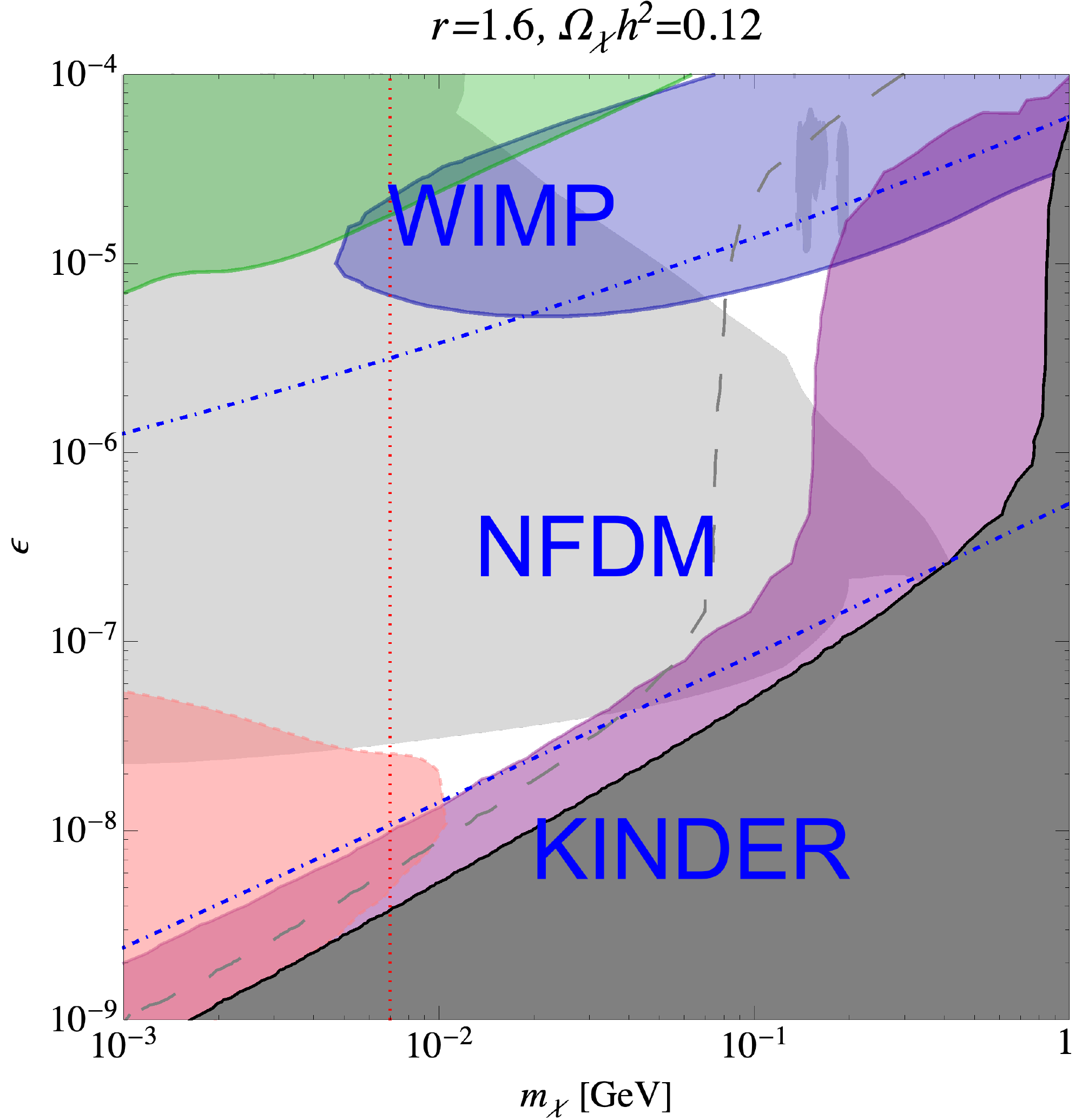}
    \includegraphics[width=0.47\textwidth]{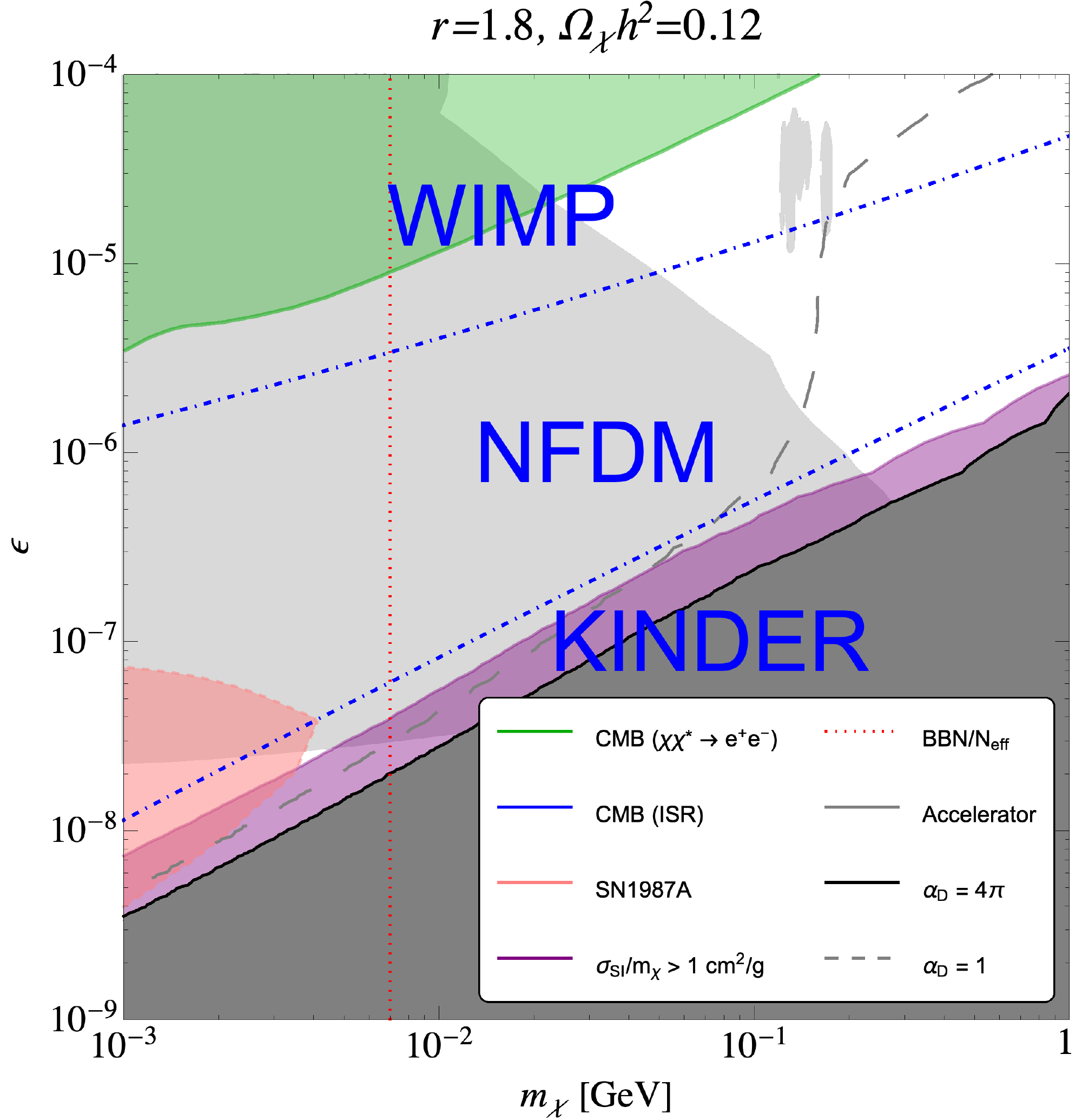}
    \caption{The $m_\chi$-$\epsilon$ parameter space for (top left) $r = 1.3$, (top right) $r = 1.4$, (bottom left) $r = 1.6$ and (bottom right) $r = 1.8$, with $\delta = 10^{-4}$. The value of $\alpha_D$ that is needed to obtain a relic abundance of $\Omega_\chi h^2 = 0.12$ has been chosen for every point on the plot. Constraints on the parameter space from the cooling of SN1987a (red), $\chi \chi \to \chi \chi$ self-interaction (purple), CMB power spectrum constraints on $\chi \chi^* \to e^+e^-$ (green) (assuming $\delta = 10^{-4}$, with constraints weakening at larger $\delta$) and $\chi \chi \to A'^* A' \to A' e^+e^-$ (blue), as well as beam experiments (light gray) are shown. A limit on electromagnetically coupled light dark matter from BBN and CMB is shown by the dotted red line; masses below the line are ruled out (assuming no other dark-sector effects). The region of the parameter space where nonperturbative values of $\alpha_D > 4 \pi$ are needed for the right relic abundance is indicated in dark gray; the dashed gray line indicates $\alpha_D = 1$. Large labels corresponding to the various regimes discussed in Ref.~\cite{Fitzpatrick:2020vba} are shown for reference. 
    } 
    \label{fig:Eps_Alpha_FDM_NFDM_Plots}
\end{figure*}

\textbf{Summary of constraints.---} Fig.~\ref{fig:Eps_Alpha_FDM_NFDM_Plots} shows the experimentally allowed regions for $r = $ 1.3, 1.4, 1.6 and 1.8 in the $m_\chi$-$\epsilon$ plane, choosing $\alpha_D$ to obtain the correct relic abundance. We also show the different freezeout phases in the $1 \lesssim r \lesssim 2$ regime, derived in Ref.~\cite{Fitzpatrick:2020vba}. In addition to some of the new constraints unique to inelastic DM discussed above, additional constraints that apply identically to the symmetric Dirac fermion case studied in Ref.~\cite{Fitzpatrick:2020vba} are shown. These include: searches for the visible decay of $A'$ at beam experiments~\cite{Bergsma:1985qz,Bergsma:1985is,Konaka:1986cb,Bjorken:1988as,Davier:1989wz,Blumlein:1990ay,Blumlein:1991xh,Gninenko:2011uv,Gninenko:2012eq,Banerjee:2016tad,Batley:2015lha,Ilten:2016tkc,Aaij:2017rft,Ilten:2018crw,Aaij:2019bvg,Tsai:2019mtm}, limits on the emission of light particles from the cooling of SN1987a~\cite{Raffelt:1987yt,Raffelt1996,Burrows1986,Burrows:1987zz,Chang:2018rso} (although arguments made in Refs.~\cite{Bar:2019ifz,Sung:2021swd} may alter or remove these limits), as well as joint BBN and CMB constraints on electromagnetically coupled DM, assuming no other dark sector effects on either of these observables~\cite{Sabti:2019mhn}. We expect direct-detection limits to be irrelevant: DM scattering with electrons and nucleons occurs only at one-loop for the ground state (with upscattering being kinematically forbidden), and is suppressed by at least an additional factor of $\epsilon^2$ relative to the symmetric scattering rate, while the scattering rate of excited states is suppressed by their tiny abundance~\cite{Zhang:2016dck,Blennow:2016gde}.

For the benchmark points where $r < 1.5$ and $r = 1.8$, a high-mass window with $\SI{30}{\mega\eV} \lesssim m_\chi \lesssim \SI{1}{\giga\eV}$ and $10^{-6} \lesssim \epsilon \lesssim 10^{-4}$ remains mostly unconstrained, with the notable exception of the LHCb search for relatively long-lived $A'$ decaying into muons~\cite{Aaij:2017rft,Aaij:2019bvg}. For $r = 1.6$, a combination of constraints from DM self-interaction and CMB constraints on DM self-annihilation with ISR closes most of the high mass window. In addition, for $r \leq 1.6$, a low-mass window near $m_\chi \sim \SI{10}{\mega\eV}$ and $\epsilon \sim 10^{-8}$ remains viable. For $r = 1.8$, the low mass window is almost completely closed due to the large dark sector couplings required for the correct relic abundance and consequently large self-interaction limits; however, this exclusion is dependent on the accuracy of the supernova constraints.  

\textbf{Conclusion.---} We have analyzed the experimental constraints on the vector-portal inelastic DM model in the regime where $1 \lesssim r \equiv m_{A'}/m_\chi \lesssim 2$, adopting the results of Ref.~\cite{Fitzpatrick:2020vba} for the relic abundance calculations in this regime. We have studied several possible DM annihilation and decay channels which are constrained by the CMB power spectrum, and have carefully examined the self-interaction constraints on this model, deriving in particular the one-loop self-scattering cross section at zero velocity for arbitrary $A'$ masses. This model has significant regions of viable parameter space that are still unexplored, providing a simple and strong motivation for a range of future experimental searches for the $A'$. The high-mass window is a potential target for beam dump experiments which are sensitive to relatively long-lived dark photons which decay visibly, including LHCb~\cite{Ilten:2015hya,Ilten:2016tkc}, FASER~\cite{Feng:2017uoz}, Belle II~\cite{Kou:2018nap}, DarkQuest/LongQuest~\cite{Berlin:2018pwi,Tsai:2019mtm}, and many other future experiments~\cite{Gninenko:2300189,Berlin:2018pwi,Adrian:2018scb,Caldwell:2018atq,Doria:2019sux,DOnofrio:2019dcp,Collaboration:2691873}. The low-mass window also motivates a more complete study of SN1987a constraints in the $1 \lesssim r \lesssim 2$ regime.

\textbf{Acknowledgments.---} The authors would like to thank Asher Berlin, Torsten Bringmann, Jae Hyeok Chang, Marco Hufnagel, Thomas G. Rizzo, Joshua Ruderman, Kai Schmidt-Hoberg, Vladyslav Shtabovenko and Yotam Soreq. This material is partially based upon PJF's work supported by the National Science Foundation Graduate Research Fellowship under Grant No. 1745302. PJF and TRS are supported by the U.S. Department of Energy, Office of Science, Office of High Energy Physics,
under grant Contract Number DE-SC0012567. HL is supported by the DOE under Award Number DESC0007968. Parts of this document were prepared by Y-DT using the resources of the Fermi National Accelerator Laboratory (Fermilab), a U.S. Department of Energy, Office of Science, HEP User Facility. Fermilab is managed by Fermi Research Alliance, LLC (FRA), acting under Contract No. DE-AC02-07CH11359. We acknowledge the use of the following software packages for this work: \texttt{FeynCalc}~\cite{Mertig:1990an,Shtabovenko:2016sxi,Shtabovenko:2020gxv}, \texttt{PackageX}~\cite{Patel:2015tea,Patel:2016fam} and \texttt{TikZ-Feynman}~\cite{Ellis:2016jkw}. We also use the compilation of dark photon bounds maintained by DarkCast~\cite{Ilten:2018crw}. Part of the work presented in this paper was performed on computational resources managed and supported by Princeton Research Computing, a consortium of groups including the Princeton Institute for Computational Science and Engineering (PICSciE) and the Office of Information Technology's High Performance Computing Center and Visualization Laboratory at Princeton University.

\bibliography{kinder}

\begin{thebibliography}{68}%
\makeatletter
\providecommand \@ifxundefined [1]{%
 \@ifx{#1\undefined}
}%
\providecommand \@ifnum [1]{%
 \ifnum #1\expandafter \@firstoftwo
 \else \expandafter \@secondoftwo
 \fi
}%
\providecommand \@ifx [1]{%
 \ifx #1\expandafter \@firstoftwo
 \else \expandafter \@secondoftwo
 \fi
}%
\providecommand \natexlab [1]{#1}%
\providecommand \enquote  [1]{``#1''}%
\providecommand \bibnamefont  [1]{#1}%
\providecommand \bibfnamefont [1]{#1}%
\providecommand \citenamefont [1]{#1}%
\providecommand \href@noop [0]{\@secondoftwo}%
\providecommand \href [0]{\begingroup \@sanitize@url \@href}%
\providecommand \@href[1]{\@@startlink{#1}\@@href}%
\providecommand \@@href[1]{\endgroup#1\@@endlink}%
\providecommand \@sanitize@url [0]{\catcode `\\12\catcode `\$12\catcode
  `\&12\catcode `\#12\catcode `\^12\catcode `\_12\catcode `\%12\relax}%
\providecommand \@@startlink[1]{}%
\providecommand \@@endlink[0]{}%
\providecommand \url  [0]{\begingroup\@sanitize@url \@url }%
\providecommand \@url [1]{\endgroup\@href {#1}{\urlprefix }}%
\providecommand \urlprefix  [0]{URL }%
\providecommand \Eprint [0]{\href }%
\providecommand \doibase [0]{http://dx.doi.org/}%
\providecommand \selectlanguage [0]{\@gobble}%
\providecommand \bibinfo  [0]{\@secondoftwo}%
\providecommand \bibfield  [0]{\@secondoftwo}%
\providecommand \translation [1]{[#1]}%
\providecommand \BibitemOpen [0]{}%
\providecommand \bibitemStop [0]{}%
\providecommand \bibitemNoStop [0]{.\EOS\space}%
\providecommand \EOS [0]{\spacefactor3000\relax}%
\providecommand \BibitemShut  [1]{\csname bibitem#1\endcsname}%
\let\auto@bib@innerbib\@empty
\bibitem [{\citenamefont {Battaglieri}\ \emph {et~al.}(2017)\citenamefont
  {Battaglieri} \emph {et~al.}}]{Battaglieri:2017aum}%
  \BibitemOpen
  \bibfield  {author} {\bibinfo {author} {\bibfnamefont {Marco}\ \bibnamefont
  {Battaglieri}} \emph {et~al.},\ }\bibfield  {title} {\enquote {\bibinfo
  {title} {{US Cosmic Visions: New Ideas in Dark Matter 2017: Community
  Report}},}\ }in\ \href@noop {} {\emph {\bibinfo {booktitle} {{U.S. Cosmic
  Visions: New Ideas in Dark Matter}}}}\ (\bibinfo {year} {2017})\ \Eprint
  {http://arxiv.org/abs/1707.04591} {arXiv:1707.04591 [hep-ph]} \BibitemShut
  {NoStop}%
\bibitem [{\citenamefont {Holdom}(1986)}]{Holdom:1985ag}%
  \BibitemOpen
  \bibfield  {author} {\bibinfo {author} {\bibfnamefont {Bob}\ \bibnamefont
  {Holdom}},\ }\bibfield  {title} {\enquote {\bibinfo {title} {{Two U(1)'s and
  Epsilon Charge Shifts}},}\ }\href {\doibase 10.1016/0370-2693(86)91377-8}
  {\bibfield  {journal} {\bibinfo  {journal} {Phys. Lett. B}\ }\textbf
  {\bibinfo {volume} {166}},\ \bibinfo {pages} {196--198} (\bibinfo {year}
  {1986})}\BibitemShut {NoStop}%
\bibitem [{\citenamefont {Pospelov}\ \emph {et~al.}(2008)\citenamefont
  {Pospelov}, \citenamefont {Ritz},\ and\ \citenamefont
  {Voloshin}}]{Pospelov:2007mp}%
  \BibitemOpen
  \bibfield  {author} {\bibinfo {author} {\bibfnamefont {Maxim}\ \bibnamefont
  {Pospelov}}, \bibinfo {author} {\bibfnamefont {Adam}\ \bibnamefont {Ritz}}, \
  and\ \bibinfo {author} {\bibfnamefont {Mikhail~B.}\ \bibnamefont
  {Voloshin}},\ }\bibfield  {title} {\enquote {\bibinfo {title} {{Secluded WIMP
  Dark Matter}},}\ }\href {\doibase 10.1016/j.physletb.2008.02.052} {\bibfield
  {journal} {\bibinfo  {journal} {Phys. Lett. B}\ }\textbf {\bibinfo {volume}
  {662}},\ \bibinfo {pages} {53--61} (\bibinfo {year} {2008})},\ \Eprint
  {http://arxiv.org/abs/0711.4866} {arXiv:0711.4866 [hep-ph]} \BibitemShut
  {NoStop}%
\bibitem [{\citenamefont {Aghanim}\ \emph {et~al.}(2020)\citenamefont {Aghanim}
  \emph {et~al.}}]{Aghanim:2018eyx}%
  \BibitemOpen
  \bibfield  {author} {\bibinfo {author} {\bibfnamefont {N.}~\bibnamefont
  {Aghanim}} \emph {et~al.} (\bibinfo {collaboration} {Planck}),\ }\bibfield
  {title} {\enquote {\bibinfo {title} {{Planck 2018 results. VI. Cosmological
  parameters}},}\ }\href {\doibase 10.1051/0004-6361/201833910} {\bibfield
  {journal} {\bibinfo  {journal} {Astron. Astrophys.}\ }\textbf {\bibinfo
  {volume} {641}},\ \bibinfo {pages} {A6} (\bibinfo {year} {2020})},\ \Eprint
  {http://arxiv.org/abs/1807.06209} {arXiv:1807.06209 [astro-ph.CO]}
  \BibitemShut {NoStop}%
\bibitem [{\citenamefont {D'Agnolo}\ and\ \citenamefont
  {Ruderman}(2015)}]{PhysRevLett.115.061301}%
  \BibitemOpen
  \bibfield  {author} {\bibinfo {author} {\bibfnamefont {Raffaele~Tito}\
  \bibnamefont {D'Agnolo}}\ and\ \bibinfo {author} {\bibfnamefont {Joshua~T.}\
  \bibnamefont {Ruderman}},\ }\bibfield  {title} {\enquote {\bibinfo {title}
  {Light dark matter from forbidden channels},}\ }\href {\doibase
  10.1103/PhysRevLett.115.061301} {\bibfield  {journal} {\bibinfo  {journal}
  {Phys. Rev. Lett.}\ }\textbf {\bibinfo {volume} {115}},\ \bibinfo {pages}
  {061301} (\bibinfo {year} {2015})}\BibitemShut {NoStop}%
\bibitem [{\citenamefont {Cline}\ \emph {et~al.}(2017)\citenamefont {Cline},
  \citenamefont {Liu}, \citenamefont {Slatyer},\ and\ \citenamefont
  {Xue}}]{Cline:2017tka}%
  \BibitemOpen
  \bibfield  {author} {\bibinfo {author} {\bibfnamefont {James~M.}\
  \bibnamefont {Cline}}, \bibinfo {author} {\bibfnamefont {Hongwan}\
  \bibnamefont {Liu}}, \bibinfo {author} {\bibfnamefont {Tracy}\ \bibnamefont
  {Slatyer}}, \ and\ \bibinfo {author} {\bibfnamefont {Wei}\ \bibnamefont
  {Xue}},\ }\bibfield  {title} {\enquote {\bibinfo {title} {{Enabling Forbidden
  Dark Matter}},}\ }\href {\doibase 10.1103/PhysRevD.96.083521} {\bibfield
  {journal} {\bibinfo  {journal} {Phys. Rev.}\ }\textbf {\bibinfo {volume}
  {D96}},\ \bibinfo {pages} {083521} (\bibinfo {year} {2017})},\ \Eprint
  {http://arxiv.org/abs/1702.07716} {arXiv:1702.07716 [hep-ph]} \BibitemShut
  {NoStop}%
\bibitem [{\citenamefont {Fitzpatrick}\ \emph {et~al.}(2020)\citenamefont
  {Fitzpatrick}, \citenamefont {Liu}, \citenamefont {Slatyer},\ and\
  \citenamefont {Tsai}}]{Fitzpatrick:2020vba}%
  \BibitemOpen
  \bibfield  {author} {\bibinfo {author} {\bibfnamefont {Patrick~J.}\
  \bibnamefont {Fitzpatrick}}, \bibinfo {author} {\bibfnamefont {Hongwan}\
  \bibnamefont {Liu}}, \bibinfo {author} {\bibfnamefont {Tracy~R.}\
  \bibnamefont {Slatyer}}, \ and\ \bibinfo {author} {\bibfnamefont {Yu-Dai}\
  \bibnamefont {Tsai}},\ }\bibfield  {title} {\enquote {\bibinfo {title} {{New
  Pathways to the Relic Abundance of Vector-Portal Dark Matter}},}\ }\href@noop
  {} {\  (\bibinfo {year} {2020})},\ \Eprint {http://arxiv.org/abs/2011.01240}
  {arXiv:2011.01240 [hep-ph]} \BibitemShut {NoStop}%
\bibitem [{\citenamefont {Tucker-Smith}\ and\ \citenamefont
  {Weiner}(2001)}]{TuckerSmith:2001hy}%
  \BibitemOpen
  \bibfield  {author} {\bibinfo {author} {\bibfnamefont {David}\ \bibnamefont
  {Tucker-Smith}}\ and\ \bibinfo {author} {\bibfnamefont {Neal}\ \bibnamefont
  {Weiner}},\ }\bibfield  {title} {\enquote {\bibinfo {title} {{Inelastic dark
  matter}},}\ }\href {\doibase 10.1103/PhysRevD.64.043502} {\bibfield
  {journal} {\bibinfo  {journal} {Phys. Rev. D}\ }\textbf {\bibinfo {volume}
  {64}},\ \bibinfo {pages} {043502} (\bibinfo {year} {2001})},\ \Eprint
  {http://arxiv.org/abs/hep-ph/0101138} {arXiv:hep-ph/0101138} \BibitemShut
  {NoStop}%
\bibitem [{\citenamefont {Finkbeiner}\ \emph {et~al.}(2011)\citenamefont
  {Finkbeiner}, \citenamefont {Goodenough}, \citenamefont {Slatyer},
  \citenamefont {Vogelsberger},\ and\ \citenamefont
  {Weiner}}]{Finkbeiner:2010sm}%
  \BibitemOpen
  \bibfield  {author} {\bibinfo {author} {\bibfnamefont {Douglas~P.}\
  \bibnamefont {Finkbeiner}}, \bibinfo {author} {\bibfnamefont {Lisa}\
  \bibnamefont {Goodenough}}, \bibinfo {author} {\bibfnamefont {Tracy~R.}\
  \bibnamefont {Slatyer}}, \bibinfo {author} {\bibfnamefont {Mark}\
  \bibnamefont {Vogelsberger}}, \ and\ \bibinfo {author} {\bibfnamefont {Neal}\
  \bibnamefont {Weiner}},\ }\bibfield  {title} {\enquote {\bibinfo {title}
  {{Consistent Scenarios for Cosmic-Ray Excesses from Sommerfeld-Enhanced Dark
  Matter Annihilation}},}\ }\href {\doibase 10.1088/1475-7516/2011/05/002}
  {\bibfield  {journal} {\bibinfo  {journal} {JCAP}\ }\textbf {\bibinfo
  {volume} {05}},\ \bibinfo {pages} {002} (\bibinfo {year} {2011})},\ \Eprint
  {http://arxiv.org/abs/1011.3082} {arXiv:1011.3082 [hep-ph]} \BibitemShut
  {NoStop}%
\bibitem [{\citenamefont {Elor}\ \emph {et~al.}(2018)\citenamefont {Elor},
  \citenamefont {Liu}, \citenamefont {Slatyer},\ and\ \citenamefont
  {Soreq}}]{Elor:2018xku}%
  \BibitemOpen
  \bibfield  {author} {\bibinfo {author} {\bibfnamefont {Gilly}\ \bibnamefont
  {Elor}}, \bibinfo {author} {\bibfnamefont {Hongwan}\ \bibnamefont {Liu}},
  \bibinfo {author} {\bibfnamefont {Tracy~R.}\ \bibnamefont {Slatyer}}, \ and\
  \bibinfo {author} {\bibfnamefont {Yotam}\ \bibnamefont {Soreq}},\ }\bibfield
  {title} {\enquote {\bibinfo {title} {{Complementarity for Dark Sector Bound
  States}},}\ }\href {\doibase 10.1103/PhysRevD.98.036015} {\bibfield
  {journal} {\bibinfo  {journal} {Phys. Rev. D}\ }\textbf {\bibinfo {volume}
  {98}},\ \bibinfo {pages} {036015} (\bibinfo {year} {2018})},\ \Eprint
  {http://arxiv.org/abs/1801.07723} {arXiv:1801.07723 [hep-ph]} \BibitemShut
  {NoStop}%
\bibitem [{\citenamefont {Baryakhtar}\ \emph {et~al.}(2020)\citenamefont
  {Baryakhtar}, \citenamefont {Berlin}, \citenamefont {Liu},\ and\
  \citenamefont {Weiner}}]{Baryakhtar:2020rwy}%
  \BibitemOpen
  \bibfield  {author} {\bibinfo {author} {\bibfnamefont {Masha}\ \bibnamefont
  {Baryakhtar}}, \bibinfo {author} {\bibfnamefont {Asher}\ \bibnamefont
  {Berlin}}, \bibinfo {author} {\bibfnamefont {Hongwan}\ \bibnamefont {Liu}}, \
  and\ \bibinfo {author} {\bibfnamefont {Neal}\ \bibnamefont {Weiner}},\
  }\bibfield  {title} {\enquote {\bibinfo {title} {{Electromagnetic Signals of
  Inelastic Dark Matter Scattering}},}\ }\href@noop {} {\  (\bibinfo {year}
  {2020})},\ \Eprint {http://arxiv.org/abs/2006.13918} {arXiv:2006.13918
  [hep-ph]} \BibitemShut {NoStop}%
\bibitem [{\citenamefont {Slatyer}(2016)}]{Slatyer:2015jla}%
  \BibitemOpen
  \bibfield  {author} {\bibinfo {author} {\bibfnamefont {Tracy~R.}\
  \bibnamefont {Slatyer}},\ }\bibfield  {title} {\enquote {\bibinfo {title}
  {{Indirect dark matter signatures in the cosmic dark ages. I. Generalizing
  the bound on s-wave dark matter annihilation from Planck results}},}\ }\href
  {\doibase 10.1103/PhysRevD.93.023527} {\bibfield  {journal} {\bibinfo
  {journal} {Phys. Rev. D}\ }\textbf {\bibinfo {volume} {93}},\ \bibinfo
  {pages} {023527} (\bibinfo {year} {2016})},\ \Eprint
  {http://arxiv.org/abs/1506.03811} {arXiv:1506.03811 [hep-ph]} \BibitemShut
  {NoStop}%
\bibitem [{\citenamefont {Poulin}\ \emph {et~al.}(2017)\citenamefont {Poulin},
  \citenamefont {Lesgourgues},\ and\ \citenamefont {Serpico}}]{Poulin:2016anj}%
  \BibitemOpen
  \bibfield  {author} {\bibinfo {author} {\bibfnamefont {Vivian}\ \bibnamefont
  {Poulin}}, \bibinfo {author} {\bibfnamefont {Julien}\ \bibnamefont
  {Lesgourgues}}, \ and\ \bibinfo {author} {\bibfnamefont {Pasquale~D.}\
  \bibnamefont {Serpico}},\ }\bibfield  {title} {\enquote {\bibinfo {title}
  {{Cosmological constraints on exotic injection of electromagnetic energy}},}\
  }\href {\doibase 10.1088/1475-7516/2017/03/043} {\bibfield  {journal}
  {\bibinfo  {journal} {JCAP}\ }\textbf {\bibinfo {volume} {03}},\ \bibinfo
  {pages} {043} (\bibinfo {year} {2017})},\ \Eprint
  {http://arxiv.org/abs/1610.10051} {arXiv:1610.10051 [astro-ph.CO]}
  \BibitemShut {NoStop}%
\bibitem [{\citenamefont {Rizzo}(2020)}]{Rizzo:2020jsm}%
  \BibitemOpen
  \bibfield  {author} {\bibinfo {author} {\bibfnamefont {Thomas~G.}\
  \bibnamefont {Rizzo}},\ }\bibfield  {title} {\enquote {\bibinfo {title}
  {{Dark Initial State Radiation and the Kinetic Mixing Portal}},}\ }\href@noop
  {} {\  (\bibinfo {year} {2020})},\ \Eprint {http://arxiv.org/abs/2006.08502}
  {arXiv:2006.08502 [hep-ph]} \BibitemShut {NoStop}%
\bibitem [{\citenamefont {Bondarenko}\ \emph {et~al.}(2021)\citenamefont
  {Bondarenko}, \citenamefont {Sokolenko}, \citenamefont {Boyarsky},
  \citenamefont {Robertson}, \citenamefont {Harvey},\ and\ \citenamefont
  {Revaz}}]{Bondarenko:2020mpf}%
  \BibitemOpen
  \bibfield  {author} {\bibinfo {author} {\bibfnamefont {Kyrylo}\ \bibnamefont
  {Bondarenko}}, \bibinfo {author} {\bibfnamefont {Anastasia}\ \bibnamefont
  {Sokolenko}}, \bibinfo {author} {\bibfnamefont {Alexey}\ \bibnamefont
  {Boyarsky}}, \bibinfo {author} {\bibfnamefont {Andrew}\ \bibnamefont
  {Robertson}}, \bibinfo {author} {\bibfnamefont {David}\ \bibnamefont
  {Harvey}}, \ and\ \bibinfo {author} {\bibfnamefont {Yves}\ \bibnamefont
  {Revaz}},\ }\bibfield  {title} {\enquote {\bibinfo {title} {{From dwarf
  galaxies to galaxy clusters: Self-Interacting Dark Matter over 7 orders of
  magnitude in halo mass}},}\ }\href {\doibase 10.1088/1475-7516/2021/01/043}
  {\bibfield  {journal} {\bibinfo  {journal} {JCAP}\ }\textbf {\bibinfo
  {volume} {01}},\ \bibinfo {pages} {043} (\bibinfo {year} {2021})},\ \Eprint
  {http://arxiv.org/abs/2006.06623} {arXiv:2006.06623 [astro-ph.CO]}
  \BibitemShut {NoStop}%
\bibitem [{\citenamefont {Schutz}\ and\ \citenamefont
  {Slatyer}(2015)}]{Schutz:2014nka}%
  \BibitemOpen
  \bibfield  {author} {\bibinfo {author} {\bibfnamefont {Katelin}\ \bibnamefont
  {Schutz}}\ and\ \bibinfo {author} {\bibfnamefont {Tracy~R.}\ \bibnamefont
  {Slatyer}},\ }\bibfield  {title} {\enquote {\bibinfo {title}
  {{Self-Scattering for Dark Matter with an Excited State}},}\ }\href {\doibase
  10.1088/1475-7516/2015/01/021} {\bibfield  {journal} {\bibinfo  {journal}
  {JCAP}\ }\textbf {\bibinfo {volume} {01}},\ \bibinfo {pages} {021} (\bibinfo
  {year} {2015})},\ \Eprint {http://arxiv.org/abs/1409.2867} {arXiv:1409.2867
  [hep-ph]} \BibitemShut {NoStop}%
\bibitem [{\citenamefont {Bergsma}\ \emph {et~al.}(1985)\citenamefont {Bergsma}
  \emph {et~al.}}]{Bergsma:1985qz}%
  \BibitemOpen
  \bibfield  {author} {\bibinfo {author} {\bibfnamefont {F.}~\bibnamefont
  {Bergsma}} \emph {et~al.} (\bibinfo {collaboration} {CHARM}),\ }\bibfield
  {title} {\enquote {\bibinfo {title} {{Search for Axion Like Particle
  Production in 400-{GeV} Proton - Copper Interactions}},}\ }\href {\doibase
  10.1016/0370-2693(85)90400-9} {\bibfield  {journal} {\bibinfo  {journal}
  {Phys. Lett.}\ }\textbf {\bibinfo {volume} {157B}},\ \bibinfo {pages}
  {458--462} (\bibinfo {year} {1985})}\BibitemShut {NoStop}%
\bibitem [{\citenamefont {Bergsma}\ \emph {et~al.}(1986)\citenamefont {Bergsma}
  \emph {et~al.}}]{Bergsma:1985is}%
  \BibitemOpen
  \bibfield  {author} {\bibinfo {author} {\bibfnamefont {F.}~\bibnamefont
  {Bergsma}} \emph {et~al.} (\bibinfo {collaboration} {CHARM}),\ }\bibfield
  {title} {\enquote {\bibinfo {title} {{A search for decays of heavy neutrinos
  in the mass range 0.5 GeV to 2.8 GeV}},}\ }\href {\doibase
  10.1016/0370-2693(86)91601-1} {\bibfield  {journal} {\bibinfo  {journal}
  {Phys. Lett.}\ }\textbf {\bibinfo {volume} {B166}},\ \bibinfo {pages} {473}
  (\bibinfo {year} {1986})}\BibitemShut {NoStop}%
\bibitem [{\citenamefont {Konaka}\ \emph {et~al.}(1986)\citenamefont {Konaka}
  \emph {et~al.}}]{Konaka:1986cb}%
  \BibitemOpen
  \bibfield  {author} {\bibinfo {author} {\bibfnamefont {A.}~\bibnamefont
  {Konaka}} \emph {et~al.},\ }\bibfield  {title} {\enquote {\bibinfo {title}
  {{Search for neutral particles in electron-beam-dump experiment}},}\
  }\bibfield  {booktitle} {\emph {\bibinfo {booktitle} {{Proceedings, 23RD
  International Conference on High Energy Physics, JULY 16-23, 1986, Berkeley,
  CA}}},\ }\href {\doibase 10.1103/PhysRevLett.57.659} {\bibfield  {journal}
  {\bibinfo  {journal} {Phys. Rev. Lett.}\ }\textbf {\bibinfo {volume} {57}},\
  \bibinfo {pages} {659} (\bibinfo {year} {1986})}\BibitemShut {NoStop}%
\bibitem [{\citenamefont {Bjorken}\ \emph {et~al.}(1988)\citenamefont
  {Bjorken}, \citenamefont {Ecklund}, \citenamefont {Nelson}, \citenamefont
  {Abashian}, \citenamefont {Church}, \citenamefont {Lu}, \citenamefont {Mo},
  \citenamefont {Nunamaker},\ and\ \citenamefont {Rassmann}}]{Bjorken:1988as}%
  \BibitemOpen
  \bibfield  {author} {\bibinfo {author} {\bibfnamefont {J.~D.}\ \bibnamefont
  {Bjorken}}, \bibinfo {author} {\bibfnamefont {S.}~\bibnamefont {Ecklund}},
  \bibinfo {author} {\bibfnamefont {W.~R.}\ \bibnamefont {Nelson}}, \bibinfo
  {author} {\bibfnamefont {A.}~\bibnamefont {Abashian}}, \bibinfo {author}
  {\bibfnamefont {C.}~\bibnamefont {Church}}, \bibinfo {author} {\bibfnamefont
  {B.}~\bibnamefont {Lu}}, \bibinfo {author} {\bibfnamefont {L.~W.}\
  \bibnamefont {Mo}}, \bibinfo {author} {\bibfnamefont {T.~A.}\ \bibnamefont
  {Nunamaker}}, \ and\ \bibinfo {author} {\bibfnamefont {P.}~\bibnamefont
  {Rassmann}},\ }\bibfield  {title} {\enquote {\bibinfo {title} {{Search for
  neutral metastable penetrating particles produced in the SLAC beam dump}},}\
  }\href {\doibase 10.1103/PhysRevD.38.3375} {\bibfield  {journal} {\bibinfo
  {journal} {Phys. Rev.}\ }\textbf {\bibinfo {volume} {D38}},\ \bibinfo {pages}
  {3375} (\bibinfo {year} {1988})}\BibitemShut {NoStop}%
\bibitem [{\citenamefont {Davier}\ and\ \citenamefont
  {Nguyen~Ngoc}(1989)}]{Davier:1989wz}%
  \BibitemOpen
  \bibfield  {author} {\bibinfo {author} {\bibfnamefont {M.}~\bibnamefont
  {Davier}}\ and\ \bibinfo {author} {\bibfnamefont {H.}~\bibnamefont
  {Nguyen~Ngoc}},\ }\bibfield  {title} {\enquote {\bibinfo {title} {{An
  unambiguous search for a light Higgs boson}},}\ }\href {\doibase
  10.1016/0370-2693(89)90174-3} {\bibfield  {journal} {\bibinfo  {journal}
  {Phys. Lett.}\ }\textbf {\bibinfo {volume} {B229}},\ \bibinfo {pages} {150}
  (\bibinfo {year} {1989})}\BibitemShut {NoStop}%
\bibitem [{\citenamefont {Bl{\"u}mlein}\ \emph {et~al.}(1991)\citenamefont
  {Bl{\"u}mlein} \emph {et~al.}}]{Blumlein:1990ay}%
  \BibitemOpen
  \bibfield  {author} {\bibinfo {author} {\bibfnamefont {J.}~\bibnamefont
  {Bl{\"u}mlein}} \emph {et~al.},\ }\bibfield  {title} {\enquote {\bibinfo
  {title} {{Limits on neutral light scalar and pseudoscalar particles in a
  proton beam dump experiment}},}\ }\href {\doibase 10.1007/BF01548556}
  {\bibfield  {journal} {\bibinfo  {journal} {Z. Phys.}\ }\textbf {\bibinfo
  {volume} {C51}},\ \bibinfo {pages} {341--350} (\bibinfo {year}
  {1991})}\BibitemShut {NoStop}%
\bibitem [{\citenamefont {Bl{\"u}mlein}\ \emph {et~al.}(1992)\citenamefont
  {Bl{\"u}mlein} \emph {et~al.}}]{Blumlein:1991xh}%
  \BibitemOpen
  \bibfield  {author} {\bibinfo {author} {\bibfnamefont {J.}~\bibnamefont
  {Bl{\"u}mlein}} \emph {et~al.},\ }\bibfield  {title} {\enquote {\bibinfo
  {title} {{Limits on the mass of light (pseudo)scalar particles from
  Bethe-Heitler $e^+ e^-$ and $\mu^+\mu^-$ pair production in a proton-iron
  beam dump experiment}},}\ }\href {\doibase 10.1142/S0217751X9200171X}
  {\bibfield  {journal} {\bibinfo  {journal} {Int. J. Mod. Phys.}\ }\textbf
  {\bibinfo {volume} {A7}},\ \bibinfo {pages} {3835--3850} (\bibinfo {year}
  {1992})}\BibitemShut {NoStop}%
\bibitem [{\citenamefont {Gninenko}(2012{\natexlab{a}})}]{Gninenko:2011uv}%
  \BibitemOpen
  \bibfield  {author} {\bibinfo {author} {\bibfnamefont {S.N.}\ \bibnamefont
  {Gninenko}},\ }\bibfield  {title} {\enquote {\bibinfo {title} {{Stringent
  limits on the $\pi^0 \to \gamma X, X \to e^+e^-$ decay from neutrino
  experiments and constraints on new light gauge bosons}},}\ }\href {\doibase
  10.1103/PhysRevD.85.055027} {\bibfield  {journal} {\bibinfo  {journal} {Phys.
  Rev.}\ }\textbf {\bibinfo {volume} {D85}},\ \bibinfo {pages} {055027}
  (\bibinfo {year} {2012}{\natexlab{a}})},\ \Eprint
  {http://arxiv.org/abs/1112.5438} {arXiv:1112.5438 [hep-ph]} \BibitemShut
  {NoStop}%
\bibitem [{\citenamefont {Gninenko}(2012{\natexlab{b}})}]{Gninenko:2012eq}%
  \BibitemOpen
  \bibfield  {author} {\bibinfo {author} {\bibfnamefont {S.N.}\ \bibnamefont
  {Gninenko}},\ }\bibfield  {title} {\enquote {\bibinfo {title} {{Constraints
  on sub-GeV hidden sector gauge bosons from a search for heavy neutrino
  decays}},}\ }\href {\doibase 10.1016/j.physletb.2012.06.002} {\bibfield
  {journal} {\bibinfo  {journal} {Phys. Lett.}\ }\textbf {\bibinfo {volume}
  {B713}},\ \bibinfo {pages} {244--248} (\bibinfo {year}
  {2012}{\natexlab{b}})},\ \Eprint {http://arxiv.org/abs/1204.3583}
  {arXiv:1204.3583 [hep-ph]} \BibitemShut {NoStop}%
\bibitem [{\citenamefont {Banerjee}\ \emph {et~al.}(2017)\citenamefont
  {Banerjee} \emph {et~al.}}]{Banerjee:2016tad}%
  \BibitemOpen
  \bibfield  {author} {\bibinfo {author} {\bibfnamefont {D.}~\bibnamefont
  {Banerjee}} \emph {et~al.} (\bibinfo {collaboration} {NA64}),\ }\bibfield
  {title} {\enquote {\bibinfo {title} {{Search for invisible decays of sub-GeV
  dark photons in missing-energy events at the CERN SPS}},}\ }\href {\doibase
  10.1103/PhysRevLett.118.011802} {\bibfield  {journal} {\bibinfo  {journal}
  {Phys. Rev. Lett.}\ }\textbf {\bibinfo {volume} {118}},\ \bibinfo {pages}
  {011802} (\bibinfo {year} {2017})},\ \Eprint
  {http://arxiv.org/abs/1610.02988} {arXiv:1610.02988 [hep-ex]} \BibitemShut
  {NoStop}%
\bibitem [{\citenamefont {Batley}\ \emph {et~al.}(2015)\citenamefont {Batley}
  \emph {et~al.}}]{Batley:2015lha}%
  \BibitemOpen
  \bibfield  {author} {\bibinfo {author} {\bibfnamefont {J.~R.}\ \bibnamefont
  {Batley}} \emph {et~al.} (\bibinfo {collaboration} {NA48/2}),\ }\bibfield
  {title} {\enquote {\bibinfo {title} {{Search for the dark photon in $\pi^0$
  decays}},}\ }\href {\doibase 10.1016/j.physletb.2015.04.068} {\bibfield
  {journal} {\bibinfo  {journal} {Phys. Lett.}\ }\textbf {\bibinfo {volume}
  {B746}},\ \bibinfo {pages} {178--185} (\bibinfo {year} {2015})},\ \Eprint
  {http://arxiv.org/abs/1504.00607} {arXiv:1504.00607 [hep-ex]} \BibitemShut
  {NoStop}%
\bibitem [{\citenamefont {Ilten}\ \emph {et~al.}(2016)\citenamefont {Ilten},
  \citenamefont {Soreq}, \citenamefont {Thaler}, \citenamefont {Williams},\
  and\ \citenamefont {Xue}}]{Ilten:2016tkc}%
  \BibitemOpen
  \bibfield  {author} {\bibinfo {author} {\bibfnamefont {Philip}\ \bibnamefont
  {Ilten}}, \bibinfo {author} {\bibfnamefont {Yotam}\ \bibnamefont {Soreq}},
  \bibinfo {author} {\bibfnamefont {Jesse}\ \bibnamefont {Thaler}}, \bibinfo
  {author} {\bibfnamefont {Mike}\ \bibnamefont {Williams}}, \ and\ \bibinfo
  {author} {\bibfnamefont {Wei}\ \bibnamefont {Xue}},\ }\bibfield  {title}
  {\enquote {\bibinfo {title} {{Proposed Inclusive Dark Photon Search at
  LHCb}},}\ }\href {\doibase 10.1103/PhysRevLett.116.251803} {\bibfield
  {journal} {\bibinfo  {journal} {Phys. Rev. Lett.}\ }\textbf {\bibinfo
  {volume} {116}},\ \bibinfo {pages} {251803} (\bibinfo {year} {2016})},\
  \Eprint {http://arxiv.org/abs/1603.08926} {arXiv:1603.08926 [hep-ph]}
  \BibitemShut {NoStop}%
\bibitem [{\citenamefont {Aaij}\ \emph {et~al.}(2017)\citenamefont {Aaij} \emph
  {et~al.}}]{Aaij:2017rft}%
  \BibitemOpen
  \bibfield  {author} {\bibinfo {author} {\bibfnamefont {Roel}\ \bibnamefont
  {Aaij}} \emph {et~al.} (\bibinfo {collaboration} {LHCb}),\ }\bibfield
  {title} {\enquote {\bibinfo {title} {{Search for dark photons produced in 13
  TeV $pp$ collisions}},}\ }\href@noop {} {\  (\bibinfo {year} {2017})},\
  \Eprint {http://arxiv.org/abs/1710.02867} {arXiv:1710.02867 [hep-ex]}
  \BibitemShut {NoStop}%
\bibitem [{\citenamefont {Ilten}\ \emph {et~al.}(2018)\citenamefont {Ilten},
  \citenamefont {Soreq}, \citenamefont {Williams},\ and\ \citenamefont
  {Xue}}]{Ilten:2018crw}%
  \BibitemOpen
  \bibfield  {author} {\bibinfo {author} {\bibfnamefont {Philip}\ \bibnamefont
  {Ilten}}, \bibinfo {author} {\bibfnamefont {Yotam}\ \bibnamefont {Soreq}},
  \bibinfo {author} {\bibfnamefont {Mike}\ \bibnamefont {Williams}}, \ and\
  \bibinfo {author} {\bibfnamefont {Wei}\ \bibnamefont {Xue}},\ }\bibfield
  {title} {\enquote {\bibinfo {title} {{Serendipity in dark photon
  searches}},}\ }\href {\doibase 10.1007/JHEP06(2018)004} {\bibfield  {journal}
  {\bibinfo  {journal} {JHEP}\ }\textbf {\bibinfo {volume} {06}},\ \bibinfo
  {pages} {004} (\bibinfo {year} {2018})},\ \Eprint
  {http://arxiv.org/abs/1801.04847} {arXiv:1801.04847 [hep-ph]} \BibitemShut
  {NoStop}%
\bibitem [{\citenamefont {Aaij}\ \emph {et~al.}(2020)\citenamefont {Aaij} \emph
  {et~al.}}]{Aaij:2019bvg}%
  \BibitemOpen
  \bibfield  {author} {\bibinfo {author} {\bibfnamefont {Roel}\ \bibnamefont
  {Aaij}} \emph {et~al.} (\bibinfo {collaboration} {LHCb}),\ }\bibfield
  {title} {\enquote {\bibinfo {title} {{Search for $A'\!\to\!\mu^+\mu^-$
  decays}},}\ }\href {\doibase 10.1103/PhysRevLett.124.041801} {\bibfield
  {journal} {\bibinfo  {journal} {Phys. Rev. Lett.}\ }\textbf {\bibinfo
  {volume} {124}},\ \bibinfo {pages} {041801} (\bibinfo {year} {2020})},\
  \Eprint {http://arxiv.org/abs/1910.06926} {arXiv:1910.06926 [hep-ex]}
  \BibitemShut {NoStop}%
\bibitem [{\citenamefont {Tsai}\ \emph {et~al.}(2019)\citenamefont {Tsai},
  \citenamefont {deNiverville},\ and\ \citenamefont {Liu}}]{Tsai:2019mtm}%
  \BibitemOpen
  \bibfield  {author} {\bibinfo {author} {\bibfnamefont {Yu-Dai}\ \bibnamefont
  {Tsai}}, \bibinfo {author} {\bibfnamefont {Patrick}\ \bibnamefont
  {deNiverville}}, \ and\ \bibinfo {author} {\bibfnamefont {Ming~Xiong}\
  \bibnamefont {Liu}},\ }\bibfield  {title} {\enquote {\bibinfo {title} {{The
  High-Energy Frontier of the Intensity Frontier: Closing the Dark Photon,
  Inelastic Dark Matter, and Muon g-2 Windows}},}\ }\href@noop {} {\  (\bibinfo
  {year} {2019})},\ \Eprint {http://arxiv.org/abs/1908.07525} {arXiv:1908.07525
  [hep-ph]} \BibitemShut {NoStop}%
\bibitem [{\citenamefont {Raffelt}\ and\ \citenamefont
  {Seckel}(1988)}]{Raffelt:1987yt}%
  \BibitemOpen
  \bibfield  {author} {\bibinfo {author} {\bibfnamefont {Georg}\ \bibnamefont
  {Raffelt}}\ and\ \bibinfo {author} {\bibfnamefont {David}\ \bibnamefont
  {Seckel}},\ }\bibfield  {title} {\enquote {\bibinfo {title} {{Bounds on
  Exotic Particle Interactions from SN 1987a}},}\ }\href {\doibase
  10.1103/PhysRevLett.60.1793} {\bibfield  {journal} {\bibinfo  {journal}
  {Phys. Rev. Lett.}\ }\textbf {\bibinfo {volume} {60}},\ \bibinfo {pages}
  {1793} (\bibinfo {year} {1988})}\BibitemShut {NoStop}%
\bibitem [{\citenamefont {Raffelt}(1996)}]{Raffelt1996}%
  \BibitemOpen
  \bibfield  {author} {\bibinfo {author} {\bibfnamefont {G.~G.}\ \bibnamefont
  {Raffelt}},\ }\href@noop {} {\emph {\bibinfo {title} {{Stars as laboratories
  for fundamental physics}}}}\ (\bibinfo  {publisher} {University of Chicago
  Press},\ \bibinfo {year} {1996})\BibitemShut {NoStop}%
\bibitem [{\citenamefont {{Burrows}}\ and\ \citenamefont
  {{Lattimer}}(1986)}]{Burrows1986}%
  \BibitemOpen
  \bibfield  {author} {\bibinfo {author} {\bibfnamefont {A.}~\bibnamefont
  {{Burrows}}}\ and\ \bibinfo {author} {\bibfnamefont {J.~M.}\ \bibnamefont
  {{Lattimer}}},\ }\bibfield  {title} {\enquote {\bibinfo {title} {{The birth
  of neutron stars}},}\ }\href {\doibase 10.1086/164405} {\bibfield  {journal}
  {\bibinfo  {journal} {\apj}\ }\textbf {\bibinfo {volume} {307}},\ \bibinfo
  {pages} {178--196} (\bibinfo {year} {1986})}\BibitemShut {NoStop}%
\bibitem [{\citenamefont {Burrows}\ and\ \citenamefont
  {Lattimer}(1987)}]{Burrows:1987zz}%
  \BibitemOpen
  \bibfield  {author} {\bibinfo {author} {\bibfnamefont {Adam}\ \bibnamefont
  {Burrows}}\ and\ \bibinfo {author} {\bibfnamefont {James~M.}\ \bibnamefont
  {Lattimer}},\ }\bibfield  {title} {\enquote {\bibinfo {title} {{Neutrinos
  from SN 1987A}},}\ }\href {\doibase 10.1086/184938} {\bibfield  {journal}
  {\bibinfo  {journal} {Astrophys. J. Lett.}\ }\textbf {\bibinfo {volume}
  {318}},\ \bibinfo {pages} {L63--L68} (\bibinfo {year} {1987})}\BibitemShut
  {NoStop}%
\bibitem [{\citenamefont {Chang}\ \emph {et~al.}(2018)\citenamefont {Chang},
  \citenamefont {Essig},\ and\ \citenamefont {McDermott}}]{Chang:2018rso}%
  \BibitemOpen
  \bibfield  {author} {\bibinfo {author} {\bibfnamefont {Jae~Hyeok}\
  \bibnamefont {Chang}}, \bibinfo {author} {\bibfnamefont {Rouven}\
  \bibnamefont {Essig}}, \ and\ \bibinfo {author} {\bibfnamefont {Samuel~D.}\
  \bibnamefont {McDermott}},\ }\bibfield  {title} {\enquote {\bibinfo {title}
  {{Supernova 1987A Constraints on Sub-GeV Dark Sectors, Millicharged
  Particles, the QCD Axion, and an Axion-like Particle}},}\ }\href {\doibase
  10.1007/JHEP09(2018)051} {\bibfield  {journal} {\bibinfo  {journal} {JHEP}\
  }\textbf {\bibinfo {volume} {09}},\ \bibinfo {pages} {051} (\bibinfo {year}
  {2018})},\ \Eprint {http://arxiv.org/abs/1803.00993} {arXiv:1803.00993
  [hep-ph]} \BibitemShut {NoStop}%
\bibitem [{\citenamefont {Bar}\ \emph {et~al.}(2020)\citenamefont {Bar},
  \citenamefont {Blum},\ and\ \citenamefont {D'Amico}}]{Bar:2019ifz}%
  \BibitemOpen
  \bibfield  {author} {\bibinfo {author} {\bibfnamefont {Nitsan}\ \bibnamefont
  {Bar}}, \bibinfo {author} {\bibfnamefont {Kfir}\ \bibnamefont {Blum}}, \ and\
  \bibinfo {author} {\bibfnamefont {Guido}\ \bibnamefont {D'Amico}},\
  }\bibfield  {title} {\enquote {\bibinfo {title} {{Is there a supernova bound
  on axions?}}}\ }\href {\doibase 10.1103/PhysRevD.101.123025} {\bibfield
  {journal} {\bibinfo  {journal} {Phys. Rev. D}\ }\textbf {\bibinfo {volume}
  {101}},\ \bibinfo {pages} {123025} (\bibinfo {year} {2020})},\ \Eprint
  {http://arxiv.org/abs/1907.05020} {arXiv:1907.05020 [hep-ph]} \BibitemShut
  {NoStop}%
\bibitem [{\citenamefont {Sung}\ \emph {et~al.}(2021)\citenamefont {Sung},
  \citenamefont {Guo},\ and\ \citenamefont {Wu}}]{Sung:2021swd}%
  \BibitemOpen
  \bibfield  {author} {\bibinfo {author} {\bibfnamefont {Allan}\ \bibnamefont
  {Sung}}, \bibinfo {author} {\bibfnamefont {Gang}\ \bibnamefont {Guo}}, \ and\
  \bibinfo {author} {\bibfnamefont {Meng-Ru}\ \bibnamefont {Wu}},\ }\bibfield
  {title} {\enquote {\bibinfo {title} {{Supernova Constraint on
  Self-Interacting Dark Sector Particles}},}\ }\href@noop {} {\  (\bibinfo
  {year} {2021})},\ \Eprint {http://arxiv.org/abs/2102.04601} {arXiv:2102.04601
  [hep-ph]} \BibitemShut {NoStop}%
\bibitem [{\citenamefont {Sabti}\ \emph {et~al.}(2020)\citenamefont {Sabti},
  \citenamefont {Alvey}, \citenamefont {Escudero}, \citenamefont {Fairbairn},\
  and\ \citenamefont {Blas}}]{Sabti:2019mhn}%
  \BibitemOpen
  \bibfield  {author} {\bibinfo {author} {\bibfnamefont {Nashwan}\ \bibnamefont
  {Sabti}}, \bibinfo {author} {\bibfnamefont {James}\ \bibnamefont {Alvey}},
  \bibinfo {author} {\bibfnamefont {Miguel}\ \bibnamefont {Escudero}}, \bibinfo
  {author} {\bibfnamefont {Malcolm}\ \bibnamefont {Fairbairn}}, \ and\ \bibinfo
  {author} {\bibfnamefont {Diego}\ \bibnamefont {Blas}},\ }\bibfield  {title}
  {\enquote {\bibinfo {title} {{Refined Bounds on MeV-scale Thermal Dark
  Sectors from BBN and the CMB}},}\ }\href {\doibase
  10.1088/1475-7516/2020/01/004} {\bibfield  {journal} {\bibinfo  {journal}
  {JCAP}\ }\textbf {\bibinfo {volume} {01}},\ \bibinfo {pages} {004} (\bibinfo
  {year} {2020})},\ \Eprint {http://arxiv.org/abs/1910.01649} {arXiv:1910.01649
  [hep-ph]} \BibitemShut {NoStop}%
\bibitem [{\citenamefont {Zhang}(2017)}]{Zhang:2016dck}%
  \BibitemOpen
  \bibfield  {author} {\bibinfo {author} {\bibfnamefont {Yue}\ \bibnamefont
  {Zhang}},\ }\bibfield  {title} {\enquote {\bibinfo {title} {{Self-interacting
  Dark Matter Without Direct Detection Constraints}},}\ }\href {\doibase
  10.1016/j.dark.2016.12.003} {\bibfield  {journal} {\bibinfo  {journal} {Phys.
  Dark Univ.}\ }\textbf {\bibinfo {volume} {15}},\ \bibinfo {pages} {82--89}
  (\bibinfo {year} {2017})},\ \Eprint {http://arxiv.org/abs/1611.03492}
  {arXiv:1611.03492 [hep-ph]} \BibitemShut {NoStop}%
\bibitem [{\citenamefont {Blennow}\ \emph {et~al.}(2017)\citenamefont
  {Blennow}, \citenamefont {Clementz},\ and\ \citenamefont
  {Herrero-Garcia}}]{Blennow:2016gde}%
  \BibitemOpen
  \bibfield  {author} {\bibinfo {author} {\bibfnamefont {Mattias}\ \bibnamefont
  {Blennow}}, \bibinfo {author} {\bibfnamefont {Stefan}\ \bibnamefont
  {Clementz}}, \ and\ \bibinfo {author} {\bibfnamefont {Juan}\ \bibnamefont
  {Herrero-Garcia}},\ }\bibfield  {title} {\enquote {\bibinfo {title}
  {{Self-interacting inelastic dark matter: A viable solution to the small
  scale structure problems}},}\ }\href {\doibase 10.1088/1475-7516/2017/03/048}
  {\bibfield  {journal} {\bibinfo  {journal} {JCAP}\ }\textbf {\bibinfo
  {volume} {03}},\ \bibinfo {pages} {048} (\bibinfo {year} {2017})},\ \Eprint
  {http://arxiv.org/abs/1612.06681} {arXiv:1612.06681 [hep-ph]} \BibitemShut
  {NoStop}%
\bibitem [{\citenamefont {Ilten}\ \emph {et~al.}(2015)\citenamefont {Ilten},
  \citenamefont {Thaler}, \citenamefont {Williams},\ and\ \citenamefont
  {Xue}}]{Ilten:2015hya}%
  \BibitemOpen
  \bibfield  {author} {\bibinfo {author} {\bibfnamefont {Philip}\ \bibnamefont
  {Ilten}}, \bibinfo {author} {\bibfnamefont {Jesse}\ \bibnamefont {Thaler}},
  \bibinfo {author} {\bibfnamefont {Mike}\ \bibnamefont {Williams}}, \ and\
  \bibinfo {author} {\bibfnamefont {Wei}\ \bibnamefont {Xue}},\ }\bibfield
  {title} {\enquote {\bibinfo {title} {{Dark photons from charm mesons at
  LHCb}},}\ }\href {\doibase 10.1103/PhysRevD.92.115017} {\bibfield  {journal}
  {\bibinfo  {journal} {Phys. Rev. D}\ }\textbf {\bibinfo {volume} {92}},\
  \bibinfo {pages} {115017} (\bibinfo {year} {2015})},\ \Eprint
  {http://arxiv.org/abs/1509.06765} {arXiv:1509.06765 [hep-ph]} \BibitemShut
  {NoStop}%
\bibitem [{\citenamefont {Feng}\ \emph {et~al.}(2018)\citenamefont {Feng},
  \citenamefont {Galon}, \citenamefont {Kling},\ and\ \citenamefont
  {Trojanowski}}]{Feng:2017uoz}%
  \BibitemOpen
  \bibfield  {author} {\bibinfo {author} {\bibfnamefont {Jonathan}\
  \bibnamefont {Feng}}, \bibinfo {author} {\bibfnamefont {Iftah}\ \bibnamefont
  {Galon}}, \bibinfo {author} {\bibfnamefont {Felix}\ \bibnamefont {Kling}}, \
  and\ \bibinfo {author} {\bibfnamefont {Sebastian}\ \bibnamefont
  {Trojanowski}},\ }\bibfield  {title} {\enquote {\bibinfo {title} {{ForwArd
  Search ExpeRiment at the LHC}},}\ }\href {\doibase
  10.1103/PhysRevD.97.035001} {\bibfield  {journal} {\bibinfo  {journal} {Phys.
  Rev.}\ }\textbf {\bibinfo {volume} {D97}},\ \bibinfo {pages} {035001}
  (\bibinfo {year} {2018})},\ \Eprint {http://arxiv.org/abs/1708.09389}
  {arXiv:1708.09389 [hep-ph]} \BibitemShut {NoStop}%
\bibitem [{\citenamefont {Altmannshofer}\ \emph {et~al.}(2019)\citenamefont
  {Altmannshofer} \emph {et~al.}}]{Kou:2018nap}%
  \BibitemOpen
  \bibfield  {author} {\bibinfo {author} {\bibfnamefont {W.}~\bibnamefont
  {Altmannshofer}} \emph {et~al.} (\bibinfo {collaboration} {Belle-II}),\
  }\bibfield  {title} {\enquote {\bibinfo {title} {{The Belle II Physics
  Book}},}\ }\href {\doibase 10.1093/ptep/ptz106} {\bibfield  {journal}
  {\bibinfo  {journal} {PTEP}\ }\textbf {\bibinfo {volume} {2019}},\ \bibinfo
  {pages} {123C01} (\bibinfo {year} {2019})},\ \bibinfo {note} {[Erratum: PTEP
  2020, 029201 (2020)]},\ \Eprint {http://arxiv.org/abs/1808.10567}
  {arXiv:1808.10567 [hep-ex]} \BibitemShut {NoStop}%
\bibitem [{\citenamefont {Berlin}\ \emph {et~al.}(2018)\citenamefont {Berlin},
  \citenamefont {Gori}, \citenamefont {Schuster},\ and\ \citenamefont
  {Toro}}]{Berlin:2018pwi}%
  \BibitemOpen
  \bibfield  {author} {\bibinfo {author} {\bibfnamefont {Asher}\ \bibnamefont
  {Berlin}}, \bibinfo {author} {\bibfnamefont {Stefania}\ \bibnamefont {Gori}},
  \bibinfo {author} {\bibfnamefont {Philip}\ \bibnamefont {Schuster}}, \ and\
  \bibinfo {author} {\bibfnamefont {Natalia}\ \bibnamefont {Toro}},\ }\bibfield
   {title} {\enquote {\bibinfo {title} {{Dark Sectors at the Fermilab SeaQuest
  Experiment}},}\ }\href {\doibase 10.1103/PhysRevD.98.035011} {\bibfield
  {journal} {\bibinfo  {journal} {Phys. Rev. D}\ }\textbf {\bibinfo {volume}
  {98}},\ \bibinfo {pages} {035011} (\bibinfo {year} {2018})},\ \Eprint
  {http://arxiv.org/abs/1804.00661} {arXiv:1804.00661 [hep-ph]} \BibitemShut
  {NoStop}%
\bibitem [{\citenamefont {Gninenko}(2018)}]{Gninenko:2300189}%
  \BibitemOpen
  \bibfield  {author} {\bibinfo {author} {\bibfnamefont {Sergei}\ \bibnamefont
  {Gninenko}},\ }\href {http://cds.cern.ch/record/2300189} {\emph {\bibinfo
  {title} {{Addendum to the NA64 Proposal: Search for the $A'\to invisible$ and
  $X\to e^+e^-$ decays in 2021}}}},\ \bibinfo {type} {Tech. Rep.}\ (\bibinfo
  {institution} {CERN},\ \bibinfo {address} {Geneva},\ \bibinfo {year}
  {2018})\BibitemShut {NoStop}%
\bibitem [{\citenamefont {Adrian}\ \emph {et~al.}(2018)\citenamefont {Adrian}
  \emph {et~al.}}]{Adrian:2018scb}%
  \BibitemOpen
  \bibfield  {author} {\bibinfo {author} {\bibfnamefont {P.~H.}\ \bibnamefont
  {Adrian}} \emph {et~al.} (\bibinfo {collaboration} {HPS}),\ }\bibfield
  {title} {\enquote {\bibinfo {title} {{Search for a dark photon in
  electroproduced $e^{+}e^{-}$ pairs with the Heavy Photon Search experiment at
  JLab}},}\ }\href {\doibase 10.1103/PhysRevD.98.091101} {\bibfield  {journal}
  {\bibinfo  {journal} {Phys. Rev. D}\ }\textbf {\bibinfo {volume} {98}},\
  \bibinfo {pages} {091101} (\bibinfo {year} {2018})},\ \Eprint
  {http://arxiv.org/abs/1807.11530} {arXiv:1807.11530 [hep-ex]} \BibitemShut
  {NoStop}%
\bibitem [{\citenamefont {Caldwell}\ \emph {et~al.}(2018)\citenamefont
  {Caldwell} \emph {et~al.}}]{Caldwell:2018atq}%
  \BibitemOpen
  \bibfield  {author} {\bibinfo {author} {\bibfnamefont {A.}~\bibnamefont
  {Caldwell}} \emph {et~al.},\ }\bibfield  {title} {\enquote {\bibinfo {title}
  {{Particle physics applications of the AWAKE acceleration scheme}},}\
  }\href@noop {} {\  (\bibinfo {year} {2018})},\ \Eprint
  {http://arxiv.org/abs/1812.11164} {arXiv:1812.11164 [physics.acc-ph]}
  \BibitemShut {NoStop}%
\bibitem [{\citenamefont {Doria}\ \emph {et~al.}(2020)\citenamefont {Doria},
  \citenamefont {Achenbach}, \citenamefont {Christmann}, \citenamefont
  {Denig},\ and\ \citenamefont {Merkel}}]{Doria:2019sux}%
  \BibitemOpen
  \bibfield  {author} {\bibinfo {author} {\bibfnamefont {Luca}\ \bibnamefont
  {Doria}}, \bibinfo {author} {\bibfnamefont {Patrick}\ \bibnamefont
  {Achenbach}}, \bibinfo {author} {\bibfnamefont {Mirco}\ \bibnamefont
  {Christmann}}, \bibinfo {author} {\bibfnamefont {Achim}\ \bibnamefont
  {Denig}}, \ and\ \bibinfo {author} {\bibfnamefont {Harald}\ \bibnamefont
  {Merkel}},\ }\bibfield  {title} {\enquote {\bibinfo {title} {{Dark Matter at
  the Intensity Frontier: the new MESA electron accelerator facility}},}\
  }\href {\doibase 10.22323/1.360.0022} {\bibfield  {journal} {\bibinfo
  {journal} {PoS}\ }\textbf {\bibinfo {volume} {ALPS2019}},\ \bibinfo {pages}
  {022} (\bibinfo {year} {2020})},\ \Eprint {http://arxiv.org/abs/1908.07921}
  {arXiv:1908.07921 [hep-ex]} \BibitemShut {NoStop}%
\bibitem [{\citenamefont {D'Onofrio}\ \emph {et~al.}(2020)\citenamefont
  {D'Onofrio}, \citenamefont {Fischer},\ and\ \citenamefont
  {Wang}}]{DOnofrio:2019dcp}%
  \BibitemOpen
  \bibfield  {author} {\bibinfo {author} {\bibfnamefont {Monica}\ \bibnamefont
  {D'Onofrio}}, \bibinfo {author} {\bibfnamefont {Oliver}\ \bibnamefont
  {Fischer}}, \ and\ \bibinfo {author} {\bibfnamefont {Zeren~Simon}\
  \bibnamefont {Wang}},\ }\bibfield  {title} {\enquote {\bibinfo {title}
  {{Searching for Dark Photons at the LHeC and FCC-he}},}\ }\href {\doibase
  10.1103/PhysRevD.101.015020} {\bibfield  {journal} {\bibinfo  {journal}
  {Phys. Rev. D}\ }\textbf {\bibinfo {volume} {101}},\ \bibinfo {pages}
  {015020} (\bibinfo {year} {2020})},\ \Eprint
  {http://arxiv.org/abs/1909.02312} {arXiv:1909.02312 [hep-ph]} \BibitemShut
  {NoStop}%
\bibitem [{\citenamefont {Collaboration}(2019)}]{Collaboration:2691873}%
  \BibitemOpen
  \bibfield  {author} {\bibinfo {author} {\bibfnamefont {NA62}\ \bibnamefont
  {Collaboration}} (\bibinfo {collaboration} {NA62 Collaboration}),\ }\href
  {https://cds.cern.ch/record/2691873} {\emph {\bibinfo {title} {{ADDENDUM I TO
  P326 Continuation of the physics programme of the NA62 experiment}}}},\
  \bibinfo {type} {Tech. Rep.}\ (\bibinfo  {institution} {CERN},\ \bibinfo
  {address} {Geneva},\ \bibinfo {year} {2019})\BibitemShut {NoStop}%
\bibitem [{\citenamefont {Mertig}\ \emph {et~al.}(1991)\citenamefont {Mertig},
  \citenamefont {Bohm},\ and\ \citenamefont {Denner}}]{Mertig:1990an}%
  \BibitemOpen
  \bibfield  {author} {\bibinfo {author} {\bibfnamefont {R.}~\bibnamefont
  {Mertig}}, \bibinfo {author} {\bibfnamefont {M.}~\bibnamefont {Bohm}}, \ and\
  \bibinfo {author} {\bibfnamefont {Ansgar}\ \bibnamefont {Denner}},\
  }\bibfield  {title} {\enquote {\bibinfo {title} {{FEYN CALC: Computer
  algebraic calculation of Feynman amplitudes}},}\ }\href {\doibase
  10.1016/0010-4655(91)90130-D} {\bibfield  {journal} {\bibinfo  {journal}
  {Comput. Phys. Commun.}\ }\textbf {\bibinfo {volume} {64}},\ \bibinfo {pages}
  {345--359} (\bibinfo {year} {1991})}\BibitemShut {NoStop}%
\bibitem [{\citenamefont {Shtabovenko}\ \emph {et~al.}(2016)\citenamefont
  {Shtabovenko}, \citenamefont {Mertig},\ and\ \citenamefont
  {Orellana}}]{Shtabovenko:2016sxi}%
  \BibitemOpen
  \bibfield  {author} {\bibinfo {author} {\bibfnamefont {Vladyslav}\
  \bibnamefont {Shtabovenko}}, \bibinfo {author} {\bibfnamefont {Rolf}\
  \bibnamefont {Mertig}}, \ and\ \bibinfo {author} {\bibfnamefont {Frederik}\
  \bibnamefont {Orellana}},\ }\bibfield  {title} {\enquote {\bibinfo {title}
  {{New Developments in FeynCalc 9.0}},}\ }\href {\doibase
  10.1016/j.cpc.2016.06.008} {\bibfield  {journal} {\bibinfo  {journal}
  {Comput. Phys. Commun.}\ }\textbf {\bibinfo {volume} {207}},\ \bibinfo
  {pages} {432--444} (\bibinfo {year} {2016})},\ \Eprint
  {http://arxiv.org/abs/1601.01167} {arXiv:1601.01167 [hep-ph]} \BibitemShut
  {NoStop}%
\bibitem [{\citenamefont {Shtabovenko}\ \emph {et~al.}(2020)\citenamefont
  {Shtabovenko}, \citenamefont {Mertig},\ and\ \citenamefont
  {Orellana}}]{Shtabovenko:2020gxv}%
  \BibitemOpen
  \bibfield  {author} {\bibinfo {author} {\bibfnamefont {Vladyslav}\
  \bibnamefont {Shtabovenko}}, \bibinfo {author} {\bibfnamefont {Rolf}\
  \bibnamefont {Mertig}}, \ and\ \bibinfo {author} {\bibfnamefont {Frederik}\
  \bibnamefont {Orellana}},\ }\bibfield  {title} {\enquote {\bibinfo {title}
  {{FeynCalc 9.3: New features and improvements}},}\ }\href {\doibase
  10.1016/j.cpc.2020.107478} {\bibfield  {journal} {\bibinfo  {journal}
  {Comput. Phys. Commun.}\ }\textbf {\bibinfo {volume} {256}},\ \bibinfo
  {pages} {107478} (\bibinfo {year} {2020})},\ \Eprint
  {http://arxiv.org/abs/2001.04407} {arXiv:2001.04407 [hep-ph]} \BibitemShut
  {NoStop}%
\bibitem [{\citenamefont {Patel}(2015)}]{Patel:2015tea}%
  \BibitemOpen
  \bibfield  {author} {\bibinfo {author} {\bibfnamefont {Hiren~H.}\
  \bibnamefont {Patel}},\ }\bibfield  {title} {\enquote {\bibinfo {title}
  {{Package-X: A Mathematica package for the analytic calculation of one-loop
  integrals}},}\ }\href {\doibase 10.1016/j.cpc.2015.08.017} {\bibfield
  {journal} {\bibinfo  {journal} {Comput. Phys. Commun.}\ }\textbf {\bibinfo
  {volume} {197}},\ \bibinfo {pages} {276--290} (\bibinfo {year} {2015})},\
  \Eprint {http://arxiv.org/abs/1503.01469} {arXiv:1503.01469 [hep-ph]}
  \BibitemShut {NoStop}%
\bibitem [{\citenamefont {Patel}(2017)}]{Patel:2016fam}%
  \BibitemOpen
  \bibfield  {author} {\bibinfo {author} {\bibfnamefont {Hiren~H.}\
  \bibnamefont {Patel}},\ }\bibfield  {title} {\enquote {\bibinfo {title}
  {{Package-X 2.0: A Mathematica package for the analytic calculation of
  one-loop integrals}},}\ }\href {\doibase 10.1016/j.cpc.2017.04.015}
  {\bibfield  {journal} {\bibinfo  {journal} {Comput. Phys. Commun.}\ }\textbf
  {\bibinfo {volume} {218}},\ \bibinfo {pages} {66--70} (\bibinfo {year}
  {2017})},\ \Eprint {http://arxiv.org/abs/1612.00009} {arXiv:1612.00009
  [hep-ph]} \BibitemShut {NoStop}%
\bibitem [{\citenamefont {Ellis}(2017)}]{Ellis:2016jkw}%
  \BibitemOpen
  \bibfield  {author} {\bibinfo {author} {\bibfnamefont {Joshua}\ \bibnamefont
  {Ellis}},\ }\bibfield  {title} {\enquote {\bibinfo {title} {{TikZ-Feynman:
  Feynman diagrams with TikZ}},}\ }\href {\doibase 10.1016/j.cpc.2016.08.019}
  {\bibfield  {journal} {\bibinfo  {journal} {Comput. Phys. Commun.}\ }\textbf
  {\bibinfo {volume} {210}},\ \bibinfo {pages} {103--123} (\bibinfo {year}
  {2017})},\ \Eprint {http://arxiv.org/abs/1601.05437} {arXiv:1601.05437
  [hep-ph]} \BibitemShut {NoStop}%
\bibitem [{\citenamefont {Berlin}\ and\ \citenamefont
  {Kling}(2019)}]{Berlin:2018jbm}%
  \BibitemOpen
  \bibfield  {author} {\bibinfo {author} {\bibfnamefont {Asher}\ \bibnamefont
  {Berlin}}\ and\ \bibinfo {author} {\bibfnamefont {Felix}\ \bibnamefont
  {Kling}},\ }\bibfield  {title} {\enquote {\bibinfo {title} {{Inelastic Dark
  Matter at the LHC Lifetime Frontier: ATLAS, CMS, LHCb, CODEX-b, FASER, and
  MATHUSLA}},}\ }\href {\doibase 10.1103/PhysRevD.99.015021} {\bibfield
  {journal} {\bibinfo  {journal} {Phys. Rev. D}\ }\textbf {\bibinfo {volume}
  {99}},\ \bibinfo {pages} {015021} (\bibinfo {year} {2019})},\ \Eprint
  {http://arxiv.org/abs/1810.01879} {arXiv:1810.01879 [hep-ph]} \BibitemShut
  {NoStop}%
\bibitem [{\citenamefont {Batell}\ \emph {et~al.}(2009)\citenamefont {Batell},
  \citenamefont {Pospelov},\ and\ \citenamefont {Ritz}}]{Batell:2009vb}%
  \BibitemOpen
  \bibfield  {author} {\bibinfo {author} {\bibfnamefont {Brian}\ \bibnamefont
  {Batell}}, \bibinfo {author} {\bibfnamefont {Maxim}\ \bibnamefont
  {Pospelov}}, \ and\ \bibinfo {author} {\bibfnamefont {Adam}\ \bibnamefont
  {Ritz}},\ }\bibfield  {title} {\enquote {\bibinfo {title} {{Direct Detection
  of Multi-component Secluded WIMPs}},}\ }\href {\doibase
  10.1103/PhysRevD.79.115019} {\bibfield  {journal} {\bibinfo  {journal} {Phys.
  Rev. D}\ }\textbf {\bibinfo {volume} {79}},\ \bibinfo {pages} {115019}
  (\bibinfo {year} {2009})},\ \Eprint {http://arxiv.org/abs/0903.3396}
  {arXiv:0903.3396 [hep-ph]} \BibitemShut {NoStop}%
\bibitem [{\citenamefont {Finkbeiner}\ \emph {et~al.}(2009)\citenamefont
  {Finkbeiner}, \citenamefont {Slatyer}, \citenamefont {Weiner},\ and\
  \citenamefont {Yavin}}]{Finkbeiner:2009mi}%
  \BibitemOpen
  \bibfield  {author} {\bibinfo {author} {\bibfnamefont {Douglas~P.}\
  \bibnamefont {Finkbeiner}}, \bibinfo {author} {\bibfnamefont {Tracy~R.}\
  \bibnamefont {Slatyer}}, \bibinfo {author} {\bibfnamefont {Neal}\
  \bibnamefont {Weiner}}, \ and\ \bibinfo {author} {\bibfnamefont {Itay}\
  \bibnamefont {Yavin}},\ }\bibfield  {title} {\enquote {\bibinfo {title}
  {{PAMELA, DAMA, INTEGRAL and Signatures of Metastable Excited WIMPs}},}\
  }\href {\doibase 10.1088/1475-7516/2009/09/037} {\bibfield  {journal}
  {\bibinfo  {journal} {JCAP}\ }\textbf {\bibinfo {volume} {09}},\ \bibinfo
  {pages} {037} (\bibinfo {year} {2009})},\ \Eprint
  {http://arxiv.org/abs/0903.1037} {arXiv:0903.1037 [hep-ph]} \BibitemShut
  {NoStop}%
\bibitem [{\citenamefont {Slatyer}\ and\ \citenamefont
  {Wu}(2017)}]{Slatyer:2016qyl}%
  \BibitemOpen
  \bibfield  {author} {\bibinfo {author} {\bibfnamefont {Tracy~R.}\
  \bibnamefont {Slatyer}}\ and\ \bibinfo {author} {\bibfnamefont {Chih-Liang}\
  \bibnamefont {Wu}},\ }\bibfield  {title} {\enquote {\bibinfo {title}
  {{General Constraints on Dark Matter Decay from the Cosmic Microwave
  Background}},}\ }\href {\doibase 10.1103/PhysRevD.95.023010} {\bibfield
  {journal} {\bibinfo  {journal} {Phys. Rev. D}\ }\textbf {\bibinfo {volume}
  {95}},\ \bibinfo {pages} {023010} (\bibinfo {year} {2017})},\ \Eprint
  {http://arxiv.org/abs/1610.06933} {arXiv:1610.06933 [astro-ph.CO]}
  \BibitemShut {NoStop}%
\bibitem [{\citenamefont {Hu}\ and\ \citenamefont {Silk}(1993)}]{Hu:1993gc}%
  \BibitemOpen
  \bibfield  {author} {\bibinfo {author} {\bibfnamefont {W.}~\bibnamefont
  {Hu}}\ and\ \bibinfo {author} {\bibfnamefont {J.}~\bibnamefont {Silk}},\
  }\bibfield  {title} {\enquote {\bibinfo {title} {{Thermalization constraints
  and spectral distortions for massive unstable relic particles}},}\ }\href
  {\doibase 10.1103/PhysRevLett.70.2661} {\bibfield  {journal} {\bibinfo
  {journal} {Phys. Rev. Lett.}\ }\textbf {\bibinfo {volume} {70}},\ \bibinfo
  {pages} {2661--2664} (\bibinfo {year} {1993})}\BibitemShut {NoStop}%
\bibitem [{\citenamefont {Ellis}\ \emph {et~al.}(1992)\citenamefont {Ellis},
  \citenamefont {Gelmini}, \citenamefont {Lopez}, \citenamefont {Nanopoulos},\
  and\ \citenamefont {Sarkar}}]{Ellis:1990nb}%
  \BibitemOpen
  \bibfield  {author} {\bibinfo {author} {\bibfnamefont {John~R.}\ \bibnamefont
  {Ellis}}, \bibinfo {author} {\bibfnamefont {G.~B.}\ \bibnamefont {Gelmini}},
  \bibinfo {author} {\bibfnamefont {Jorge~L.}\ \bibnamefont {Lopez}}, \bibinfo
  {author} {\bibfnamefont {Dimitri~V.}\ \bibnamefont {Nanopoulos}}, \ and\
  \bibinfo {author} {\bibfnamefont {Subir}\ \bibnamefont {Sarkar}},\ }\bibfield
   {title} {\enquote {\bibinfo {title} {{Astrophysical constraints on massive
  unstable neutral relic particles}},}\ }\href {\doibase
  10.1016/0550-3213(92)90438-H} {\bibfield  {journal} {\bibinfo  {journal}
  {Nucl. Phys. B}\ }\textbf {\bibinfo {volume} {373}},\ \bibinfo {pages}
  {399--437} (\bibinfo {year} {1992})}\BibitemShut {NoStop}%
\bibitem [{\citenamefont {Chluba}(2013)}]{Chluba:2013wsa}%
  \BibitemOpen
  \bibfield  {author} {\bibinfo {author} {\bibfnamefont {Jens}\ \bibnamefont
  {Chluba}},\ }\bibfield  {title} {\enquote {\bibinfo {title} {{Distinguishing
  different scenarios of early energy release with spectral distortions of the
  cosmic microwave background}},}\ }\href {\doibase 10.1093/mnras/stt1733}
  {\bibfield  {journal} {\bibinfo  {journal} {Mon. Not. Roy. Astron. Soc.}\
  }\textbf {\bibinfo {volume} {436}},\ \bibinfo {pages} {2232--2243} (\bibinfo
  {year} {2013})},\ \Eprint {http://arxiv.org/abs/1304.6121} {arXiv:1304.6121
  [astro-ph.CO]} \BibitemShut {NoStop}%
\bibitem [{\citenamefont {Chluba}\ and\ \citenamefont
  {Jeong}(2014)}]{Chluba:2013pya}%
  \BibitemOpen
  \bibfield  {author} {\bibinfo {author} {\bibfnamefont {Jens}\ \bibnamefont
  {Chluba}}\ and\ \bibinfo {author} {\bibfnamefont {Donghui}\ \bibnamefont
  {Jeong}},\ }\bibfield  {title} {\enquote {\bibinfo {title} {{Teasing bits of
  information out of the CMB energy spectrum}},}\ }\href {\doibase
  10.1093/mnras/stt2327} {\bibfield  {journal} {\bibinfo  {journal} {Mon. Not.
  Roy. Astron. Soc.}\ }\textbf {\bibinfo {volume} {438}},\ \bibinfo {pages}
  {2065--2082} (\bibinfo {year} {2014})},\ \Eprint
  {http://arxiv.org/abs/1306.5751} {arXiv:1306.5751 [astro-ph.CO]} \BibitemShut
  {NoStop}%
\bibitem [{\citenamefont {Denner}\ \emph
  {et~al.}(1992{\natexlab{a}})\citenamefont {Denner}, \citenamefont {Eck},
  \citenamefont {Hahn},\ and\ \citenamefont {Kublbeck}}]{Denner:1992me}%
  \BibitemOpen
  \bibfield  {author} {\bibinfo {author} {\bibfnamefont {Ansgar}\ \bibnamefont
  {Denner}}, \bibinfo {author} {\bibfnamefont {H.}~\bibnamefont {Eck}},
  \bibinfo {author} {\bibfnamefont {O.}~\bibnamefont {Hahn}}, \ and\ \bibinfo
  {author} {\bibfnamefont {J.}~\bibnamefont {Kublbeck}},\ }\bibfield  {title}
  {\enquote {\bibinfo {title} {{Compact Feynman rules for Majorana
  fermions}},}\ }\href {\doibase 10.1016/0370-2693(92)91045-B} {\bibfield
  {journal} {\bibinfo  {journal} {Phys. Lett. B}\ }\textbf {\bibinfo {volume}
  {291}},\ \bibinfo {pages} {278--280} (\bibinfo {year}
  {1992}{\natexlab{a}})}\BibitemShut {NoStop}%
\bibitem [{\citenamefont {Denner}\ \emph
  {et~al.}(1992{\natexlab{b}})\citenamefont {Denner}, \citenamefont {Eck},
  \citenamefont {Hahn},\ and\ \citenamefont {Kublbeck}}]{Denner:1992vza}%
  \BibitemOpen
  \bibfield  {author} {\bibinfo {author} {\bibfnamefont {Ansgar}\ \bibnamefont
  {Denner}}, \bibinfo {author} {\bibfnamefont {H.}~\bibnamefont {Eck}},
  \bibinfo {author} {\bibfnamefont {O.}~\bibnamefont {Hahn}}, \ and\ \bibinfo
  {author} {\bibfnamefont {J.}~\bibnamefont {Kublbeck}},\ }\bibfield  {title}
  {\enquote {\bibinfo {title} {{Feynman rules for fermion number violating
  interactions}},}\ }\href {\doibase 10.1016/0550-3213(92)90169-C} {\bibfield
  {journal} {\bibinfo  {journal} {Nucl. Phys. B}\ }\textbf {\bibinfo {volume}
  {387}},\ \bibinfo {pages} {467--481} (\bibinfo {year}
  {1992}{\natexlab{b}})}\BibitemShut {NoStop}%
\end{thebibliography}%

\clearpage

\onecolumngrid
\appendix

\begin{center}
\textbf{\large New Thermal Relic Targets for Inelastic Vector-Portal Dark Matter} \\ 
\vspace{0.05in}
{\it \large Supplemental Material}\\ 
\vspace{0.05in}
{Patrick J. Fitzpatrick, Hongwan Liu, Tracy R. Slatyer, and Yu-Dai Tsai}
\end{center}

\section{Inelastic Dark Matter Models and Dark Higgs Self-Interaction}

In this section, we provide two dark sector vector-portal models that lead to the inelastic DM model described in the main text, after dark U(1) symmetry breaking. The different symmetry breaking patterns in each model lead to different self-interaction strengths through a dark Higgs exchange, and a careful examination is required to ensure that self-interaction through an exchange of a dark Higgs can be neglected, as claimed in the main Letter. 

\subsection{Models} 

At high energies, the dark sector has a U(1) gauge symmetry with an associated gauge boson $A'$, a single Dirac fermion $\Psi$ charged under the U(1) gauge group and a Dirac mass $m_D$, as well as a dark Higgs field $\Phi$, which also carries a dark U(1) charge. The dark sector $A'$ kinetically mixes with the SM photon with kinetic mixing parameter $\epsilon$. The important terms in the Lagrangian describing this model are
\begin{alignat}{1}
    \mathcal{L} \supset -\frac{1}{4} F'_{\mu\nu} F'^{\mu\nu} - \frac{\epsilon}{2 \cos \theta_w} F'_{\mu\nu}B^{\mu\nu} + \overline{\Psi} \left(i\gamma^\mu(\partial_\mu + i g_D A'_\mu) - m_D\right)\Psi + \left| D_\mu \Phi \right|^2 + V(\Phi) + \mathcal{L}_{\Phi \Psi} \,.
\end{alignat}
Here, $B^{\mu\nu}$ is the U(1)$_\text{Y}$ hypercharge field tensor, which gives the gauge invariant kinetic mixing term at high energies, and the kinetic mixing is normalized so that after electroweak symmetry breaking, the mixing becomes $(\epsilon/2)F'_{\mu\nu}F^{\mu\nu}$ ($\theta_w$ is the weak mixing angle). $V(\Phi)$ is the potential in the Higgs sector, which we assume to be of a suitable form to break the U(1) symmetry prior to the start of freezeout, with the dark Higgs particle $h_D$ having a mass $m_{h_D}$ heavier than the other dark sector particles. The covariant derivative $D_\mu \equiv \partial_\mu + i g_D Q_\Phi A_\mu$ depends on the charge of $Q_\Phi$, which we will choose appropriately below. Finally, the term $\mathcal{L}_{\Phi \Psi}$ contains interaction terms between $\Phi$ and $\Psi$. 

With an appropriate choice of $m_D$, $Q_\Phi$ and $\mathcal{L}_{\Phi \Psi}$, the symmetry breaking of the U(1) symmetry also splits the Dirac fermion  into two Majorana fermions with slightly different masses. We consider two example choices that have been studied in the literature:
\begin{enumerate}
    \item Large $m_D$, $Q_\Phi = 1$ and $\mathcal{L}_{\Phi \Psi} = (1/2)(y_D / \Lambda) (\overline{\Psi^C} \Psi \Phi^* \Phi^*$ + \text{h.c.}), a dimension-5 interaction term suppressed by the scale $\Lambda$, between $\Psi$ and $\Phi$~\cite{Finkbeiner:2010sm};

    \item Small $m_D$, $Q_\Phi = 2$ and $\mathcal{L}_{\Phi \Psi} = y_D (\overline{\Psi^C} \Psi \Phi^* + \text{h.c.})$, a Yukawa interaction with a charge-2 $\Phi$-field is introduced instead~\cite{Elor:2018xku}. 
\end{enumerate} 
In both cases, working in unitary gauge where after symmetry breaking, we have $\Phi \to (v_D + h_D)/\sqrt{2}$, we see that a Majorana mass term $(m_M/2) (\overline{\Psi^C} \Psi + \text{h.c.})$ is generated, with $m_M = y_D v_D^2 / (2 \Lambda)$ in Model~(1), and $m_M = \sqrt{2} v_D y_D$ in Model~(2). In Model~(1), $m_M$ is parametrically suppressed by a very large scale, and naturally gives a small Majorana mass compared to the Dirac mass, i.e., $m_M \ll m_D$: the dark matter particle is a pseudo-Dirac fermion. Conversely, in the second model, $m_M$ can be as large as $v_D$, leading naturally to a large Majorana mass; in this case, we will choose $m_D \ll m_M$ instead.

In either model, we can now diagonalize the fermion mass matrix to obtain the mass eigenstates. Writing $\Psi$ as a pair of Weyl fermions $(\xi, \eta^\dagger)$, the fermion mass terms can be written as
\begin{alignat}{1}
    \frac{1}{2} \begin{pmatrix}
        \xi & \eta^\dagger
    \end{pmatrix} \begin{pmatrix}
        m_M & m_D \\
        m_D & m_M 
    \end{pmatrix} \begin{pmatrix}
        \xi \\ \eta^\dagger
    \end{pmatrix} + \text{h.c.}
\end{alignat}
For Model~(1), with $m_D \gg m_M$, the mass eigenstates are $\chi^* = (\eta + \xi)/\sqrt{2}$ with mass $m_M + m_D$, and $\chi = i(\eta - \xi)/\sqrt{2}$ with mass $m_\chi \equiv |m_M - m_D|$. The splitting between the two states is $m_\chi \delta = y_D v_D^2 / \Lambda$. In Model~(2), with $m_M \gg m_D$, the mass eigenstates are instead $\chi^* = (\eta + \xi)/\sqrt{2}$ with mass $m_M + m_D$, and $\chi = (\eta - \xi)/\sqrt{2}$ with mass $m_\chi \equiv m_M - m_D$. Here, the splitting between the two states is $m_\chi \delta = 2 m_D$. 

With a slight abuse of notation, we can define four-component Majorana spinors $\chi$ and $\chi^*$, which are simply constructed from their respective, previously defined two-component versions. The coupling to the dark photon can then be simply written as
\begin{alignat}{1}
    g_D A_\mu' \overline{\Psi} \gamma^\mu \Psi = g_D A_\mu' \overline{\chi^*} \gamma^\mu \chi \,,
\end{alignat}
giving the off-diagonal interaction in the part of the Lagrangian shown in the main Letter. 

The interaction terms between the dark Higgs and the DM states depend on the details of the model. In Model~(1), we have
\begin{alignat}{1}
    \mathcal{L}_{\Phi \Psi} \supset \frac{y_D v_D}{2 \Lambda} \left(\overline{\chi^*} \chi^* h_D - \overline{\chi} \chi h_D\right) + \frac{y_D}{4 \Lambda} \left(\overline{\chi^*} \chi^* h_D^2 - \overline{\chi} \chi h_D^2\right) \,.
\end{alignat}
On the other hand, for Model~(2), we obtain 
\begin{alignat}{1}
    \mathcal{L}_{\Phi \Psi} \supset \frac{y_D}{\sqrt{2}} \left(\overline{\chi^*} \chi^* h_D + \overline{\chi} \chi h_D \right) \,.
\end{alignat}

Finally, the vev of the dark Higgs breaks the U(1) symmetry and gives the dark photon a mass of $m_{A'} = Q_\Phi g_D v_D$. We can therefore re-express $v_D$ in terms of the phenomenologically important parameters for DM freezeout and detection. In both models, we can also express $g_{h_D}$, the coupling constant between $\chi$, $\chi^*$ and $h_D$, in terms of these parameters. 

In Model~(1), $m_\chi$ is a free parameter, set approximately by the bare Dirac mass in the Lagrangian.  The splitting on the other hand is given by $\delta m_\chi = y_D m_{A'}^2/(2 g_D^2 \Lambda)$, which can be suitably adjusted by choosing $y_D/\Lambda$ appropriately. Once the splitting is chosen, however, $g_{h_D}$ is fixed for constant $g_D$ and $r$, since $g_{h_D} = y_D m_{A'} / (2 g_D \Lambda) = g_D \delta / r$. 

In Model~(2), we find that $m_\chi \approx y_D m_{A'} / (\sqrt{2} g_D)$, while the splitting is a free parameter set by the bare Dirac mass. The expression for $m_\chi$ also sets $g_{h_D} = y_D \approx \sqrt{2} g_D / r$. Unlike Model~(1), where the dark Higgs interactions originate from an irrelevant operator and are necessarily suppressed by a small parameter, there is no natural suppression on the coupling in Model~(2). 

\subsection{Dark Higgs Self-Interaction}

The difference in the coupling constant between $\chi$, $\chi^*$ and $h_D$ between these two models leads to slightly different expressions for the dark-Higgs-mediated self-interaction rate between $\chi$ particles. We will now derive the scattering cross section in each model, and discuss the conditions under which dark Higgs self-interactions can be neglected, as is done in the Letter.

In the limit of low velocity, the self-interaction scattering cross section via an exchange of a dark Higgs is
\begin{alignat}{1}
    \sigma = \frac{g_{h_D}^4}{8 \pi} \frac{m_\chi^2}{m_{h_D}^4} \,,
\end{alignat}
where $m_{h_D}$ is the mass of $h_D$. For Model~(1), we have $g_{h_D} = g_D \delta / r$, and so the scattering cross section takes the form: 
\begin{align}
    \sigma_{(1)} & = 2 \pi \alpha_D^2 \left(\frac{\delta}{r}\right)^4 \frac{m_\chi^2}{m_{h_D}^4} \,.
\end{align}  
We see that in this model the elastic scattering cross-section due to Higgs exchange is parametrically suppressed by $(\delta/r)^4$, as well as by $1/m_{h_D}^4$, since the mass splitting controls the size of the Yukawa coupling. Consequently we expect this rate to be very subdominant for $\delta \ll 10^{-1}$, even if $m_{h_D}$ is not much larger than $m_\chi$.

For Model~(2), we have $g_{h_D} = \sqrt{2} g_D / r$, and we find instead
\begin{alignat}{1}
    \sigma_{(2)} = 32\pi \frac{\alpha_D^2}{r^4} \frac{m_\chi^2}{m_{h_D}^4} \,.
\end{alignat}
We see that in this case the self-interaction cross-section is expected to be parametrically of order $\alpha_D^2/m_\chi^2$, unless $m_{h_D} \gg m_\chi$, and so may dominate over the 1-loop elastic scattering cross section, depending on $\alpha_D$ and $m_{h_D}$. For simplicity, we neglect all dark Higgs self-interaction diagrams, either by adopting Model~(1), or by assuming that $m_{h_D}$ is large enough to neglect such interactions in Model~(2). 

\section{Additional Constraints on Dark Matter Energy Injection}

We discuss two main sources of energy injection from our dark sector that may face additional constraints: excited state decays, and the one-loop annihilation of $\chi \chi \to \ell^+\ell^-$, where $\ell$ is either an electron or a muon. We will first discuss each process in turn, and then end the section by showing that these processes have limits that are subleading to those discussed in the main Letter, or are model dependent. 

\subsection{Excited State Decays}

Due to the kinetic mixing between $A'$ and Standard Model particles, the excited state can decay through the emission of an off-shell $A'$ into various Standard Model states. Depending on the size of the splitting between ground and excited states, different channels are kinematically allowed, and result in important differences in the primordial abundance of excited states. 

If $m_\chi \delta > 2 m_e$, the excited state can decay into an electron/positron pair. The lifetime in the limit where $2m_e / m_\chi \ll \delta \ll 1$ is:
\begin{alignat}{1}
    \tau_{\chi^* \to \chi e^+ e^-} &\simeq \frac{15 \pi m_{A'}^4}{4 \epsilon^2 \alpha_\text{EM} \alpha_D m_\chi^5 \delta^5} = \SI{7.0e-5}{\second} \left(\frac{r}{1.6}\right)^4 \left(\frac{10^{-3}}{\epsilon}\right)^2 \left(\frac{1.0}{\alpha_D}\right) \left(\frac{\SI{1}{\giga\eV}}{m_\chi}\right) \left(\frac{10^{-2}}{\delta}\right)^5 \,.
\end{alignat}
With this decay channel open, any primordial population of $\chi^*$ produced after the freezeout of $\chi^* \chi^* \to \chi \chi$ is depleted on the timescale of $\tau_{\chi^* \to \chi e^+e^-}$, leading to a negligible primordial population of $\chi^*$. The lack of any primordial $\chi^*$ removes the CMB constraint on $\chi^* \chi \to e^+e^-$ discussed in the main text, while the other constraints mentioned in the main text remain unchanged. However, in order to avoid constraints from energy injection during the BBN epoch, the decay lifetime must be sufficiently short (roughly less than 1 second), which can be achieved if $\delta$ varies with $m_\chi$ appropriately. While allowing $\chi^* \to \chi e^+e^-$ is likely possible for this model, we choose $m_\chi \delta < 2 m_e$ for the sake of simplicity, allowing us to fix a value of $\delta$ for all $m_\chi$ with ease.

For $m_\chi \delta < 2 m_e$, the allowed decay processes are significantly more suppressed. The kinetic mixing term between $A'$ and photons is achieved in a UV-complete manner through the following kinetic mixing term above the electroweak symmetry breaking scale (see e.g., Ref.~\cite{Berlin:2018jbm}): 
\begin{alignat}{1}
    \mathcal{L} \supset -\frac{\epsilon}{2 \cos \theta_w} F_{\mu\nu}' B^{\mu\nu} \,, 
\end{alignat}
where $B_{\mu\nu}$ is the field strength tensor of the U(1) hypercharge gauge boson. After electroweak symmetry breaking, this term generates both the kinetic mixing term between $A'$ and photons discussed in the main text and a mixing between $A'$ and the $Z$-boson. This mixing allows the decay of the excited state into the ground state and two neutrinos, $\chi^* \to \chi \nu \overline{\nu}$. We calculated the decay width in the mass basis using the results of Ref.~\cite{Berlin:2018jbm}, obtaining (see also Refs.~\cite{Batell:2009vb,Finkbeiner:2009mi} for similar calculations)
\begin{alignat}{1}
    \tau_{\chi^* \to \chi \nu \overline{\nu}} \simeq \frac{945 \pi m_{A'}^4 m_Z^4 \cos^4 \theta_w}{4 \epsilon^2 \alpha_\text{EM} \alpha_D \delta^9 m_\chi^9} = \SI{1.8e27}{\second} \left(\frac{r}{1.6}\right)^4 \left(\frac{10^{-3}}{\epsilon}\right)^2 \left(\frac{1.0}{\alpha_D}\right) \left(\frac{\SI{100}{\mega\eV}}{m_\chi}\right)^5 \left(\frac{10^{-3}}{\delta}\right)^9 \,.
    \label{eq:chi*_to_chi_nu_nu}
\end{alignat}

Another possible decay channel is $\chi^* \to \chi + 3 \gamma$, again through an off-shell photon $A'$ and an electron loop. Following Ref.~\cite{Batell:2009vb}, we estimate this decay lifetime in the limit where $\delta m_\chi \lesssim m_e$ to be
\begin{alignat}{1}
    \tau_{\chi^* \to \chi + 3\gamma} \sim \frac{2^7 3^6 5^3 \pi^3 m_e^8}{17 \alpha_\text{EM}^3 \alpha_D \epsilon^2 \delta^9 m_\chi^9} \frac{m_{A'}^4}{\alpha_D \delta^4 m_\chi^4} = \SI{1.1e18}{\second} \left(\frac{r}{1.6}\right)^4 \left(\frac{10^{-3}}{\epsilon}\right)^2 \left(\frac{1.0}{\alpha_D}\right)^2 \left(\frac{\SI{100}{\mega\eV}}{m_\chi}\right)^9 \left(\frac{10^{-3}}{\delta}\right)^{13} \,.
    \label{eq:chi*_to_chi_3_photons}
\end{alignat}
\textbf{}Eqs.~\eqref{eq:chi*_to_chi_nu_nu} and~\eqref{eq:chi*_to_chi_3_photons} show that the lifetime of $\chi^*$ is generically much longer than the age of the Universe for $\delta \leq 10^{-3}$, and so the primordially produced $\chi^*$ population is not significantly depleted by such processes. 

\subsection{One-Loop Ground-State Annihilation}

The ground state $\chi$ can annihilate to a pair of fermions at the one-loop level. We have verified that in the limit of zero $\chi$ momentum and zero electron mass, the one-loop amplitude vanishes, indicating that the velocity-averaged annihilation cross section $\langle \sigma v \rangle_{\chi \chi \to f \overline{f}}$ is either $p$-wave suppressed or helicity suppressed (or possibly both, leading to an even larger suppression). We can therefore write $\langle \sigma v \rangle$ parametrically as 
\begin{alignat}{1}
    \langle \sigma v \rangle_{\chi \chi \to f \overline{f}} \sim \frac{g_D^4 \epsilon^4 e^4}{16 \pi^2} \frac{h(m_f^2/m_\chi^2, v^2)}{8 \pi m_\chi^2} \,,
    \label{eq:one_loop_chi_chi_ann}
\end{alignat}
where $h(m_f^2/m_\chi^2, v^2)$ is a suppression factor of either $v^2$ or $m_f^2/m_\chi^2$. The first factor comprises the coupling constants in the loop diagram, as well as a loop factor. The factor of $(8 \pi m_\chi^2)^{-1}$ is simply the phase space available to the final states~\cite{Cline:2017tka}.

\begin{figure*}[t!]
    \centering
    \includegraphics[width=0.47\textwidth]{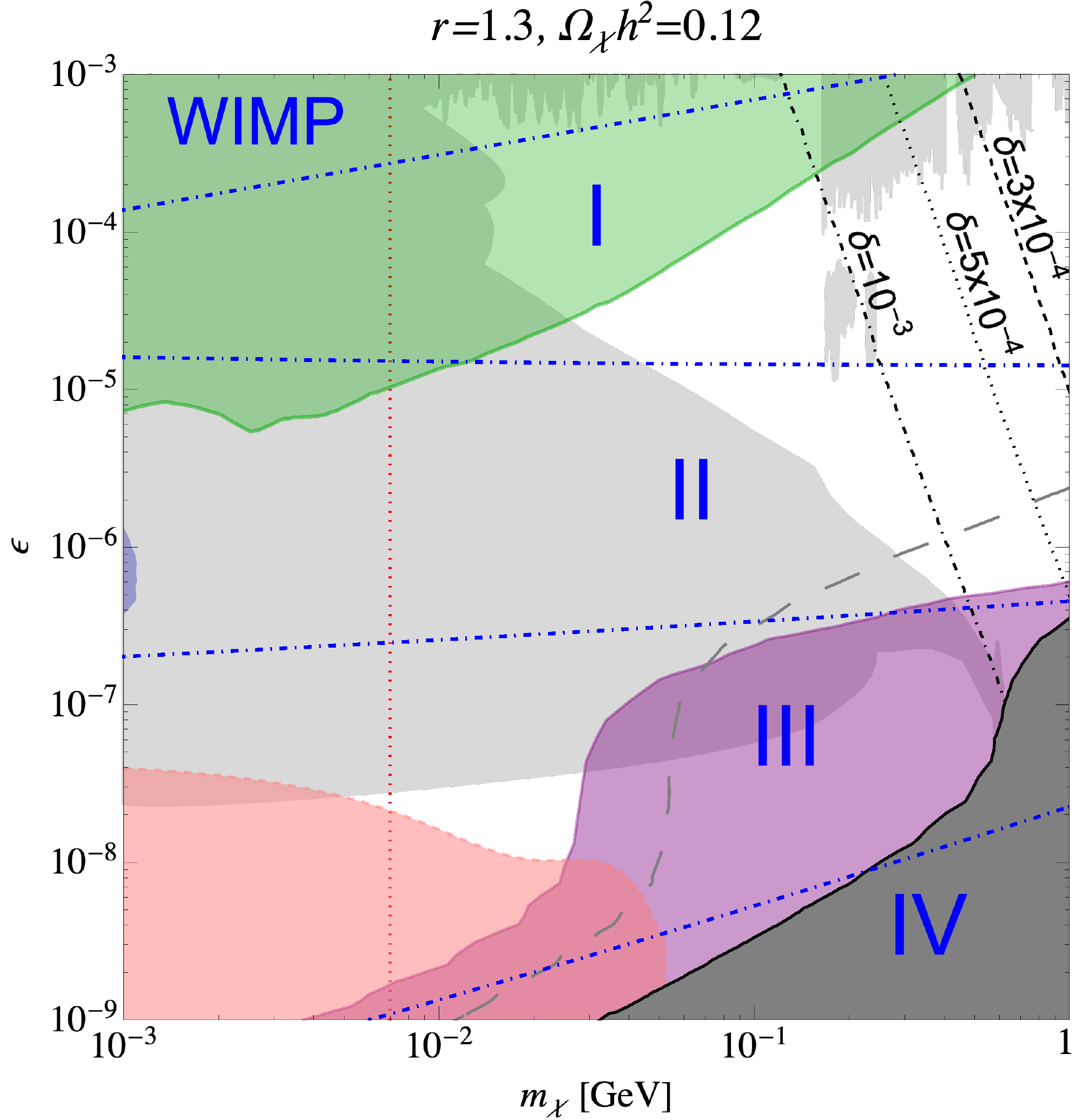} \qquad
    \includegraphics[width=0.47\textwidth]{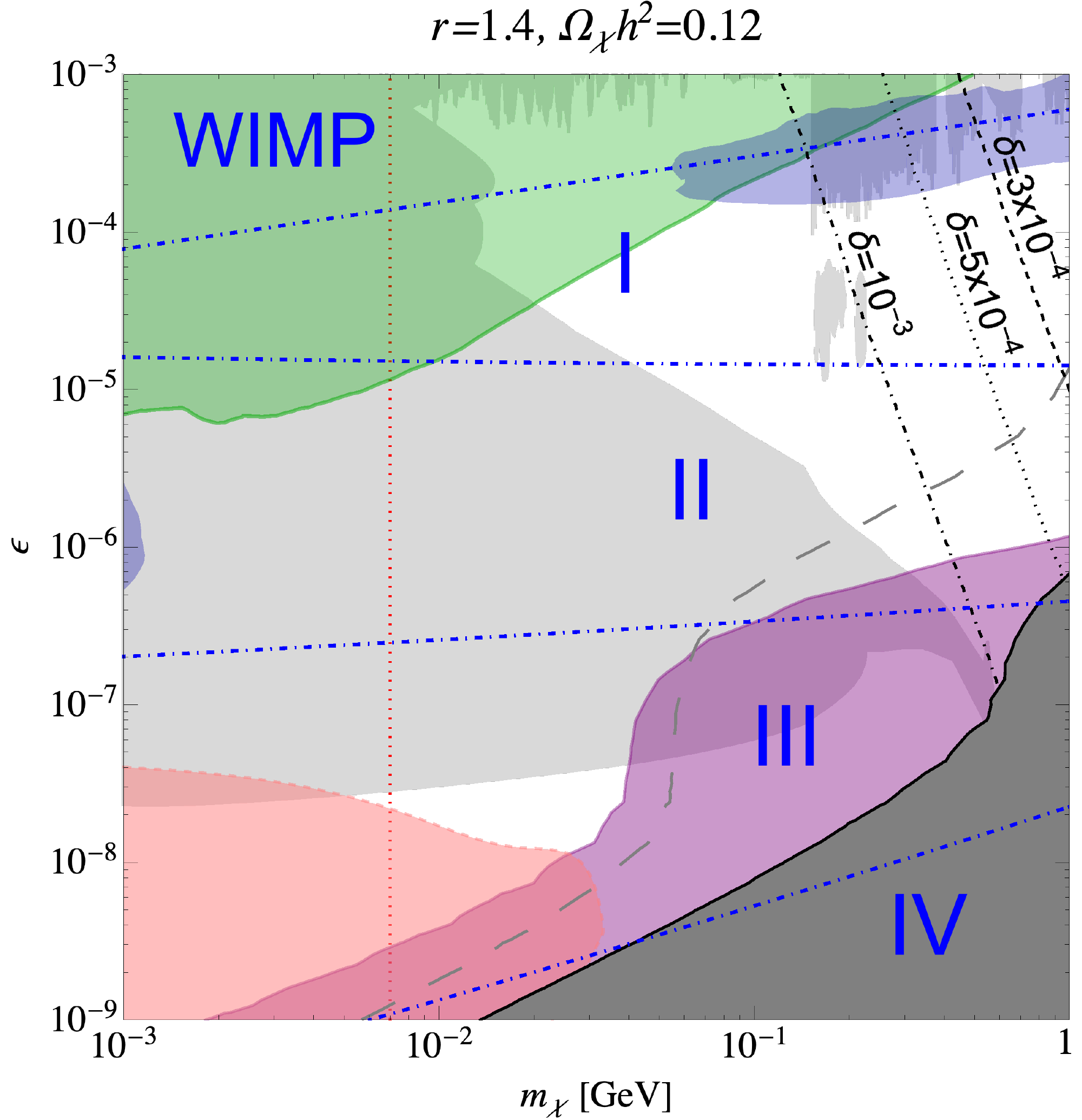} \\
    \includegraphics[width=0.47\textwidth]{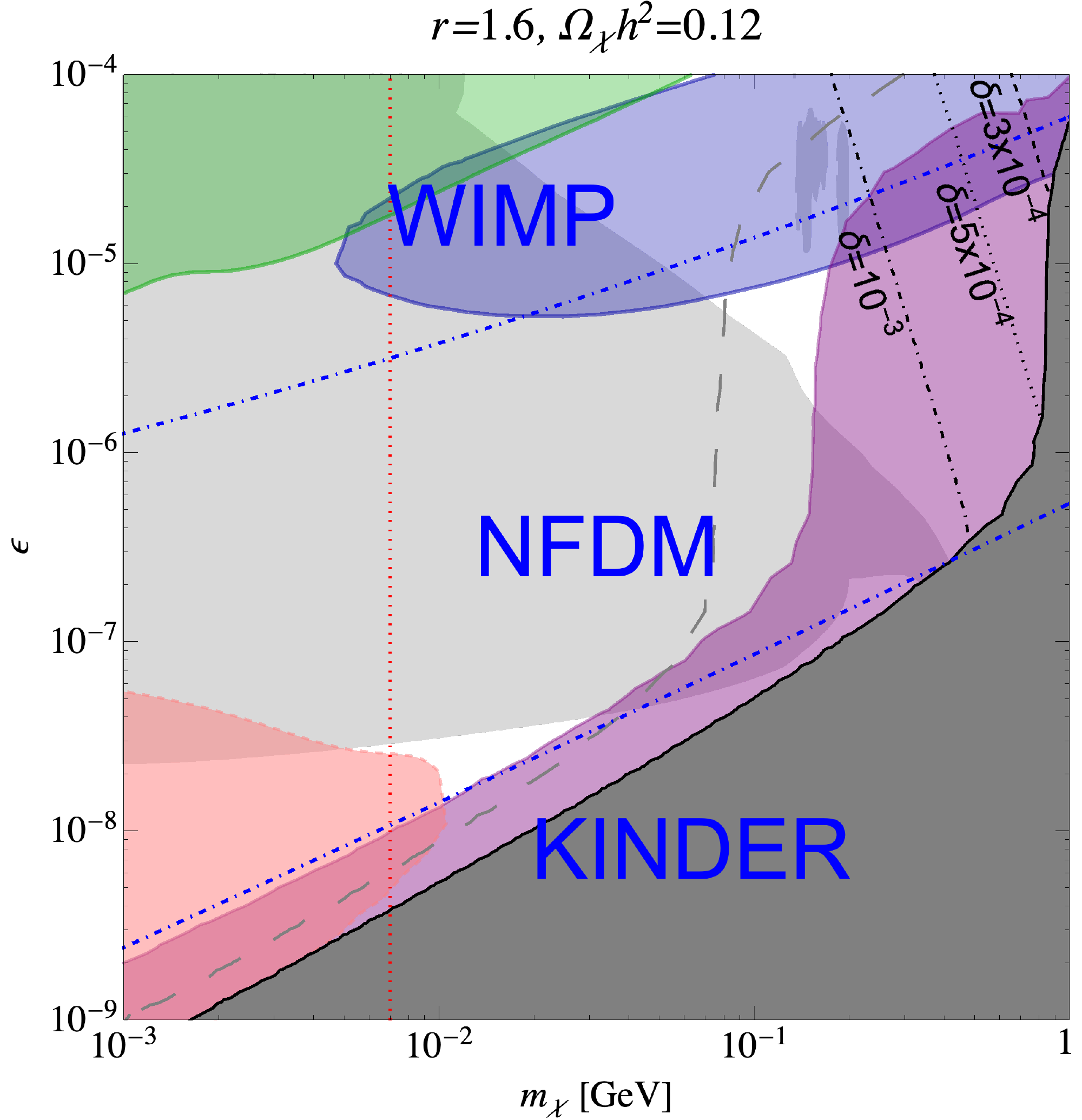} \qquad
    \includegraphics[width=0.47\textwidth]{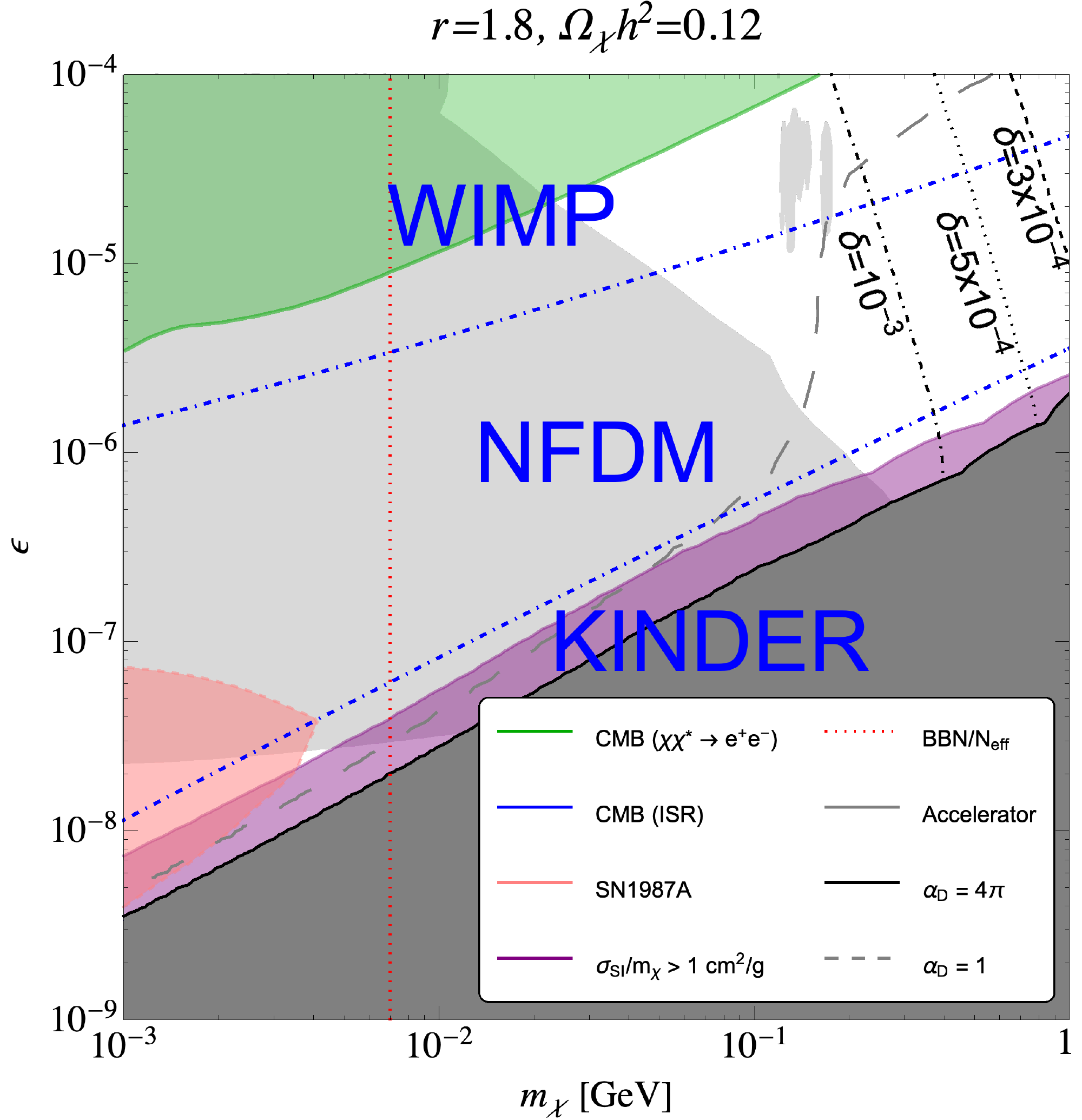}
    \caption{Lines indicating CMB power spectrum constraints on $\chi^* \to \chi + 3 \gamma$ for (black dashed) $\delta = 3\times 10^{-4} $, (black dotted) $\delta = 5\times 10^{-4} $ and (black dot-dashed) $\delta = 10^{-3} $ for (top left) $r = 1.3$, (top right) $r = 1.4$, (bottom left) $r = 1.6$ and (bottom right) $r = 1.8$ (regions above the lines are ruled out). The value of $\alpha_D$ that is needed to obtain a relic abundance of $\Omega_\chi h^2 = 0.12$ has been chosen for every point on the plot. Constraints on the parameter space from the cooling of SN1987a (red), $\chi \chi \to \chi \chi$ self-interaction (purple), CMB power spectrum constraints on $\chi \chi^* \to e^+e^-$ (green) and $\chi \chi \to A'^* A' \to A' e^+e^-$ (blue), as well as beam experiments (light gray) are shown. A limit on electromagnetically coupled light dark matter from BBN and CMB is shown by the dotted red line; masses below the line are ruled out (assuming no other dark-sector effects). Regions of the parameter space where nonperturbative values of $\alpha_D > 4 \pi$ are needed for the right relic abundance is indicated in dark gray. The dashed gray line indicates $\alpha_D = 1$. Large labels corresponding to the various regimes discussed in Ref.~\cite{Fitzpatrick:2020vba} are shown for reference.} 
    \label{fig:3_phot_decay_plots}
\end{figure*}

\subsection{Additional Constraints}

The excited state decay $\chi^* \to \chi + 3 \gamma$ leads to the injection of high-energy photons into the Universe during the cosmic dark ages; such processes are constrained by the CMB power spectrum, since they can increase the ionization fraction during the cosmic dark ages~\cite{Slatyer:2016qyl}. We can estimate the constraints on the $\chi^*$ lifetime from these results as
\begin{alignat}{1}
    \tau_{\chi^* \to \chi + 3 \gamma} > \SI{e25}{\second} \left(\frac{n_{\chi^*}}{n_\chi} \right) \,. 
    \label{eq:limit_on_tau_chi_star_to_chi_plus_3_phot}
\end{alignat}
In Fig.~\ref{fig:3_phot_decay_plots}, we show the region of parameter space ruled out by this requirement. Since the decay lifetime is extremely sensitive to $\delta$, we can see that the constraints vary rapidly as $\delta$ changes over an order of magnitude. However, the constraints are mostly confined to large values of $m_\chi$ and $\epsilon$, leaving the overall viability of this model unchanged even for $\delta \sim 10^{-3}$; there are no constraints from this process for $\delta \leq 10^{-4}$. 

Note that Eq.~\eqref{eq:limit_on_tau_chi_star_to_chi_plus_3_phot} is not strictly correct: CMB power spectrum constraints weaken significantly once the decay lifetime is shorter than the age of the Universe at recombination. Limits from CMB spectral distortion~\cite{Hu:1993gc,Ellis:1990nb,Chluba:2013wsa,Chluba:2013pya} and Big Bang Nucleosynthesis~\cite{Poulin:2016anj} become relevant, but are much less sensitive to scenarios where only a small subcomponent of DM decays, releasing energy that is much lower than its rest mass. We neglect this possibility in this discussion for simplicity, since it would simply remove the constraints at large values of $\epsilon$ and $m_\chi$, with no real impact on the overall viability of the model.

The other process that may present additional constraints is the one-loop annihilation $\chi \chi \to f \overline{f}$, with cross section given in Eq.~\eqref{eq:one_loop_chi_chi_ann}, which may be constrained by the CMB power spectrum limits on DM annihilation. To check if such processes are important, we compare the one-loop rate with the $\chi \chi^* \to f \overline{f}$ annihilation process considered in the main text. This sets the following limit for when the one-loop $\chi \chi \to f \overline{f}$ is important: 
\begin{alignat}{1}
    \alpha_D \alpha_\text{EM} \epsilon^2 h(m_f^2/m_\chi^2, v^2) > \frac{n_{\chi^*}}{n_\chi} \,.
\end{alignat}
Across our entire parameter space, this requirement is never met regardless of the form of $h$, and thus $\chi \chi \to f \overline{f}$ can be safely ignored in favor of the other constraints examined in the main Letter.

\section{One-Loop Self-Interaction Cross Section}

\begin{figure}[t!]
    \centering
    \begin{tikzpicture}
        \begin{feynman}
            \vertex(i1) {\(\chi\)}; 
            \vertex[right=1.5cm of i1] (a);
            \vertex[right=1.5cm of a]  (b);
            \vertex[right=1.5cm of b]  (f1) {\(\chi\)};

            \vertex[below=2cm of i1] (i2) {\(\chi\)};
            \vertex[right=1.5cm of i2] (c);
            \vertex[right=1.5cm of c]  (d);
            \vertex[right=1.5cm of d]  (f2) {\(\chi\)};

            \diagram*{
                (i1) -- (a) -- [edge label=\(\chi^*\)] (b),
                (b) -- (f1),
                (i2) -- (c) -- [edge label'=\(\chi^*\)] (d),
                (d) -- (f2),
                (a) -- [boson, edge label'=\(A'\)] (c),
                (b) -- [boson, edge label=\(A'\)] (d),
            };
        \end{feynman}
    \end{tikzpicture}
    \begin{tikzpicture}
        \begin{feynman}
            \vertex(i1) {\(\chi\)}; 
            \vertex[right=1.5cm of i1] (a);
            \vertex[right=1.5cm of a]  (b);
            \vertex[right=1.5cm of b]  (f1) {\(\chi\)};

            \vertex[below=2cm of i1] (i2) {\(\chi\)};
            \vertex[right=1.5cm of i2] (c);
            \vertex[right=1.5cm of c]  (d);
            \vertex[right=1.5cm of d]  (f2) {\(\chi\)};

            \diagram*{
                (i1) -- (a) -- [edge label=\(\chi^*\)] (b),
                (b) -- (f1),
                (i2) -- (c) -- [edge label'=\(\chi^*\)] (d),
                (d) -- (f2),
                (a) -- [boson, edge label'=\(A'\,\,\,\,\)] (d),
                (b) -- [boson, edge label=\(\,\,\,\,\,A'\)] (c),
            };
        \end{feynman}
    \end{tikzpicture}
    \begin{tikzpicture}
        \begin{feynman}
            \vertex(i1) {\(\chi\)}; 
            \vertex[right=1.5cm of i1] (a);
            \vertex[right=1.5cm of a]  (b);
            \vertex[right=1.5cm of b]  (f1) {\(\chi\)};

            \vertex[below=2cm of i1] (i2) {\(\chi\)};
            \vertex[right=1.5cm of i2] (c);
            \vertex[right=1.5cm of c]  (d);
            \vertex[right=1.5cm of d]  (f2) {\(\chi\)};

            \diagram*{
                (i1) -- (a), 
                (a) -- [edge label'=\(\chi^*\)] (c),
                (c) -- (i2),
                (f1) -- (b),
                (b) -- [edge label=\(\chi^*\)] (d),
                (d) -- (f2),
                (a) -- [boson, edge label=\(A'\)] (b),
                (c) -- [boson, edge label'=\(A'\)] (d),
            };
        \end{feynman}
    \end{tikzpicture}
    \caption{One-loop Feynman diagrams for the self-interaction $\chi \chi \to \chi \chi$ scattering. There are three more diagrams related to these three by interchanging the final fermionic states, giving a total of six diagrams. We follow the conventions of Refs.~\cite{Denner:1992me,Denner:1992vza} and do not put arrows on Majorana fermion propagators and external states.}
    \label{fig:self_interaction_feynman_diagrams}
\end{figure}
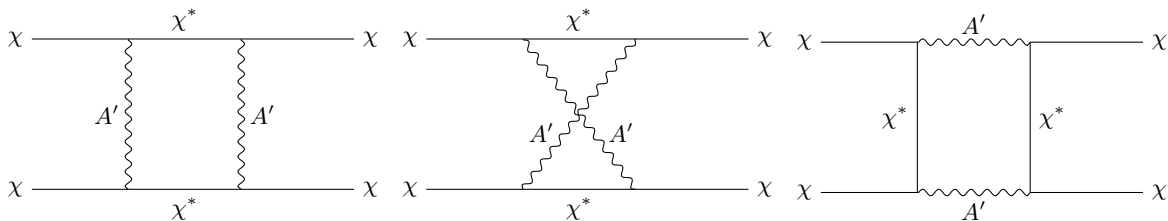

Neglecting dark Higgs self-scattering, the self-interaction scattering process $\chi \chi \to \chi \chi$ occurs at lowest order at the one-loop level, since the $A'$ couples off-diagonally to $\chi$ and $\chi^*$. There are six Feynman diagrams contributing to the amplitude of this process; three of these diagrams are shown in Fig.~\ref{fig:self_interaction_feynman_diagrams}, while the remaining three are related to these diagrams by interchanging the fermionic final states. We follow Refs.~\cite{Denner:1992me,Denner:1992vza} and their treatment of Majorana fermions and their interactions. 

We compute the one-loop self-interaction cross section $\sigma$ in the limit of zero momentum for the particles; this makes the computation of the one-loop diagrams more tractable, and is also a reasonable assumption, since we expect finite momentum corrections on the order of $v \sim 10^{-3}$ in typical dark matter structures today. We also neglect the splitting between the ground and excited states for the excited states in the loop, which would produce order $\delta$ corrections. We use \texttt{FeynCalc}~\cite{Mertig:1990an,Shtabovenko:2016sxi,Shtabovenko:2020gxv} for symbolic manipulation of our amplitudes, and \texttt{PackageX}~\cite{Patel:2015tea,Patel:2016fam} for the loop integral itself. Ultimately, we find
\begin{alignat}{1}
    \sigma = \frac{\overline{|\mathcal{M}|^2}}{128 \pi m_\chi^2} \,,
\end{alignat}
where $\overline{|\mathcal{M}|^2}$ is the squared amplitude, averaged over initial spin states and summed over final spin states. The full expressions for $\overline{|\mathcal{M}|^2}$ in four different regimes of $r$ are given in Eqs.~\eqref{eq:self_interaction_Msq_r_lt_1}--\eqref{eq:self_interaction_Msq_r_gt_2}, located at the end of this appendix. 
\begin{figure}
    \centering
    \includegraphics[width=0.5\textwidth]{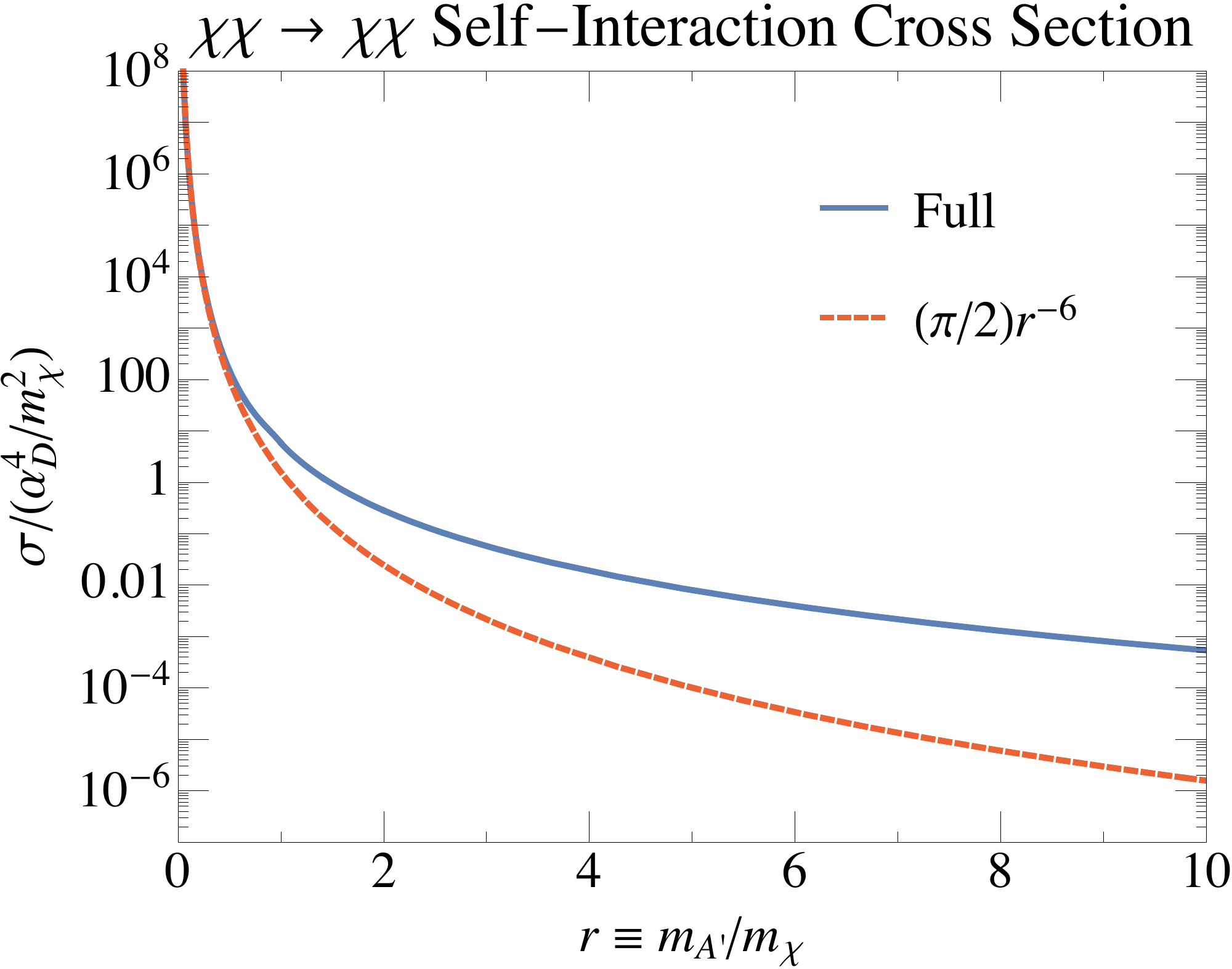}
    \caption{Self-interaction cross section $\sigma$ of $\chi \chi \to \chi \chi$ at zero momentum in units of $\alpha_D^4/m_\chi^2$ as a function of $r \equiv m_{A'}/m_\chi$ (blue). We also show the second Born approximation result at zero momentum computed assuming $r \ll 1$ in Ref.~\cite{Schutz:2014nka} for comparison (red, dashed), taking into account identical particles in the initial and final fermionic states.}
    \label{fig:xsecplot}
\end{figure}
Fig.~\ref{fig:xsecplot} shows $\sigma$ in units of $\alpha_D^4/m_\chi^2$, as a function of $r$. Although the analytic expressions for $\overline{|\mathcal{M}|^2}$ have to be written in different forms for different $r$ regimes, the function is actually smooth over all values of $r$. For comparison, we include the second Born approximation of $\sigma$ at zero momentum computed using nonrelativistic quantum mechanics techniques in Ref.~\cite{Schutz:2014nka} in the limit where $r \ll 1$, which found a cross section of $\sigma = (\pi / 2)r^{-6}$, neglecting $\delta$ and applying a correction factor to account for identical particles in the initial and final states (Ref.~\cite{Schutz:2014nka} implicitly assumed distinguishable particles). The one-loop computation performed here is equivalent to a second Born approximation of $\sigma$ at zero momentum, and for $r \ll 1$, we find
\begin{alignat}{1}
    \lim_{r \to 0} \sigma(r) \simeq \frac{\alpha_D^4}{m_\chi^2} \left[ \frac{\pi}{2r^6} + \frac{7\pi}{8 r^4} - \frac{\log(2r) + 5/3}{r^3} + \frac{23 \pi}{16 r^2} + \mathcal{O}(r^{-1}) \right]\,,
\end{alignat}
in excellent agreement with the nonrelativistic quantum mechanical result; in the range of interest to this paper, $1 \lesssim r \lesssim 2$, we obtain a cross section that is 3 -- 12 times larger than the $r \to 0$ limit would predict. For $r \gg 1$, we find
\begin{alignat}{1}
    \lim_{r \to \infty} \sigma(r) \simeq \frac{\alpha_D^4}{m_\chi^2} \left[ \frac{18}{\pi r^4} + \frac{118 - 96 \log r}{\pi r^6} + \mathcal{O}(r^{-8}) \right] \,.
\end{alignat}
The full expressions for $\overline{|\mathcal{M}|^2}$ are as follows: 

\begin{enumerate}
    \item $r < 1$:
    \begin{multline}
        \frac{\overline{\left| \mathcal{M} \right|^2}}{256 \pi^4 \alpha_D^4} = -\frac{\log ^2\left(-r^2-2 \sqrt{1-r^2}+2\right) \left(r^2-1\right)^3}{\pi ^4 \left(r^2-2\right)^4}
        +\left(\frac{p(r)}{\pi ^4 \left(r^2-2\right)^4}-\frac{4 \sqrt{1-r^2} \left(r^2-1\right)}{\pi ^4 \left(r^2-2\right)^2}\right) \log ^2(r) \\
        +\frac{16 r^2 \left(r^2-4\right) \left(r^2-2\right)^2 \left(r^2-1\right)^2-\pi ^2 k(r)}{4 \pi ^4 r^6 \left(r^2-4\right)^3 \left(r^2-2\right)^2} \\{}
        +\left[\frac{f(r)^2}{\pi ^3 r^6 \left(r^2-4\right)^3 \left(r^2-2\right)^4}+\left(\frac{4 g(r)\sqrt{1-r^2} \sqrt{4-r^2}}{\pi ^4 r^3 \left(r^2-4\right)^2 \left(r^2-2\right)^4}-\frac{2 f(r) \sqrt{4-r^2}}{\pi ^4 r^3 \left(r^2-4\right)^2 \left(r^2-2\right)^2}\right) \log (r) \right. \\
        \left.+\frac{4 g(r)\sqrt{4-r^2}}{\pi ^4 r^5 \left(r^2-4\right)^3 \left(r^2-2\right)^2}\right] \arctan \left(\frac{r}{\sqrt{4-r^2}}\right) \\
        +\left[-\frac{4 \left(r^2-1\right)}{\pi ^4 r^2 \left(r^2-4\right)}+\left(\frac{8 \left(r^2-1\right)^2}{\pi ^4 r^2 \left(r^2-4\right) \left(r^2-2\right)^2}-\frac{2 g(r)\sqrt{4-r^2}}{\pi ^3 r^3 \left(r^2-4\right)^2 \left(r^2-2\right)^4}\right) \sqrt{1-r^2} \right. \\
        \left. +\frac{f(r) \sqrt{4-r^2}}{\pi ^3 r^3 \left(r^2-4\right)^2 \left(r^2-2\right)^2}\right] \log (r) \\
        +\log \left(-r^2-2 \sqrt{1-r^2}+2\right) \left[\left(\frac{4 \left(r^2-1\right)^3}{\pi ^4 \left(r^2-2\right)^4}+\frac{2 \sqrt{1-r^2} \left(r^2-1\right)}{\pi ^4 \left(r^2-2\right)^2}\right) \log (r) \right. \\
        \left. +\left(\frac{g(r)\sqrt{4-r^2}}{\pi ^3 r^3 \left(r^2-4\right)^2 \left(r^2-2\right)^4}-\frac{4 \left(r^2-1\right)^2}{\pi ^4 r^2 \left(r^2-4\right) \left(r^2-2\right)^2}\right) \sqrt{1-r^2} \right. \\
        \left. -\frac{2 f(r) \sqrt{1-r^2} \sqrt{4-r^2} \left(r^2-1\right) \arctan \left(\frac{r}{\sqrt{4-r^2}}\right)}{\pi ^4 r^3 \left(r^2-4\right)^2 \left(r^2-2\right)^4}\right]-\frac{2 g(r)\sqrt{4-r^2}}{\pi ^3 r^5 \left(r^2-4\right)^3 \left(r^2-2\right)^2}-\frac{f(r)^2 \arctan^2\left(\frac{r}{\sqrt{4-r^2}}\right)}{\pi ^4 r^6 \left(r^2-4\right)^3 \left(r^2-2\right)^4} \,.
        \label{eq:self_interaction_Msq_r_lt_1}
    \end{multline}
    
    \item $1 \leq r < \sqrt{2}$:
    
    \begin{multline}
        \frac{\overline{\left| \mathcal{M} \right|^2}}{256 \pi^4 \alpha_D^4}  = \arctan\left(\frac{r}{\sqrt{4-r^2}}\right) \left[ \frac{f(r)^2}{\pi ^3 r^6 \left(r^2-4\right)^3 \left(r^2-2\right)^4}-\frac{2 f(r) \sqrt{4-r^2} \log (r)}{\pi ^4 r^3 \left(r^2-4\right)^2 \left(r^2-2\right)^2} \right. \\
        \left. -\frac{2 f(r) \sqrt{4-r^2} \left(r^2-1\right)^{3/2} \arctan \left(\frac{2 \sqrt{r^2-1}}{2-r^2}\right)}{\pi ^4 r^3 \left(r^2-4\right)^2 \left(r^2-2\right)^4}+\frac{4 g(r) \sqrt{4-r^2}}{\pi ^4 r^5 \left(r^2-4\right)^3 \left(r^2-2\right)^2}+\frac{2 g(r) \sqrt{4-r^2} \sqrt{r^2-1}}{\pi ^3 r^3 \left(r^2-4\right)^2 \left(r^2-2\right)^4}\right] \\
        +\left(\frac{f(r) \sqrt{4-r^2}}{\pi ^3 r^3 \left(r^2-4\right)^2 \left(r^2-2\right)^2}-\frac{2 \left(r^2-1\right)^{3/2}}{\pi ^3 \left(r^2-2\right)^2}-\frac{4 \left(r^2-1\right)}{\pi ^4 r^2 \left(r^2-4\right)}\right) \log (r) \\
        -\frac{2 g(r) \sqrt{4-r^2}}{\pi ^3 r^5 \left(r^2-4\right)^3 \left(r^2-2\right)^2}+\sqrt{r^2-1} \left(\frac{4 \left(r^2-1\right)^2}{\pi ^3 r^2 \left(r^2-4\right) \left(r^2-2\right)^2}-\frac{g(r) \sqrt{4-r^2}}{\pi ^2 r^3 \left(r^2-4\right)^2 \left(r^2-2\right)^4}\right) \\
        +\arctan\left(\frac{2 \sqrt{r^2-1}}{2-r^2}\right) \left[\sqrt{r^2-1} \left(\frac{g(r) \sqrt{4-r^2}}{\pi ^3 r^3 \left(r^2-4\right)^2 \left(r^2-2\right)^4}-\frac{4 \left(r^2-1\right)^2}{\pi ^4 r^2 \left(r^2-4\right) \left(r^2-2\right)^2}\right) \right. \\
        \left. -\frac{2 \left(r^2-1\right)^3}{\pi ^3 \left(r^2-2\right)^4}+\frac{2 \left(r^2-1\right)^{3/2} \log (r)}{\pi ^4 \left(r^2-2\right)^2}\right]+\frac{16 r^2 \left(r^2-4\right) \left(r^2-1\right)^2-\frac{\pi ^2 q(r)}{\left(r^2-2\right)^4}}{4 \pi ^4 r^6 \left(r^2-4\right)^3} \\
        -\frac{f(r)^2 \arctan^2\left(\frac{r}{\sqrt{4-r^2}}\right)}{\pi ^4 r^6 \left(r^2-4\right)^3 \left(r^2-2\right)^4} + \frac{\left(r^2-1\right)^3 \arctan^2\left(\frac{2 \sqrt{r^2-1}}{2-r^2}\right)}{\pi ^4 \left(r^2-2\right)^4}+\frac{\log ^2(r)}{\pi ^4} \,.
        \label{eq:self_interaction_Msq_1_lt_r_lt_sqrt_2}
    \end{multline}
    
    \item $\sqrt{2} \leq r < 2$:
    \begin{multline}
        \frac{\overline{\left| \mathcal{M} \right|^2}}{256 \pi^4 \alpha_D^4}  = \arctan\left(\frac{r}{\sqrt{4-r^2}}\right) \left[\frac{f(r)^2}{\pi ^3 r^6 \left(r^2-4\right)^3 \left(r^2-2\right)^4}-\frac{2 f(r) \sqrt{4-r^2} \log (r)}{\pi ^4 r^3 \left(r^2-4\right)^2 \left(r^2-2\right)^2} \right. \\
        \left. -\frac{2 f(r) \sqrt{4-r^2} \left(r^2-1\right)^{3/2} \arctan \left(\frac{2 \sqrt{r^2-1}}{2-r^2}\right)}{\pi ^4 r^3 \left(r^2-4\right)^2 \left(r^2-2\right)^4}+\frac{4 g(r) \sqrt{4-r^2}}{\pi ^4 r^5 \left(r^2-4\right)^3 \left(r^2-2\right)^2}\right] \\
        +\left(\frac{f(r) \sqrt{4-r^2}}{\pi ^3 r^3 \left(r^2-4\right)^2 \left(r^2-2\right)^2}-\frac{4 \left(r^2-1\right)}{\pi ^4 r^2 \left(r^2-4\right)}\right) \log (r) -\frac{2 g(r) \sqrt{4-r^2}}{\pi ^3 r^5 \left(r^2-4\right)^3 \left(r^2-2\right)^2}\\
        +\arctan\left(\frac{2 \sqrt{r^2-1}}{2-r^2}\right) \left[\sqrt{r^2-1} \left(\frac{g(r) \sqrt{4-r^2}}{\pi ^3 r^3 \left(r^2-4\right)^2 \left(r^2-2\right)^4}-\frac{4 \left(r^2-1\right)^2}{\pi ^4 r^2 \left(r^2-4\right) \left(r^2-2\right)^2}\right) \right. \\
        \left. +\frac{2 \left(r^2-1\right)^{3/2} \log (r)}{\pi ^4 \left(r^2-2\right)^2}\right] +\frac{16 r^2 \left(r^2-4\right) \left(r^2-1\right)^2-\frac{\pi ^2 f(r)^2}{\left(r^2-2\right)^4}}{4 \pi ^4 r^6 \left(r^2-4\right)^3} \\
        - \frac{f(r)^2 \arctan^2 \left(\frac{r}{\sqrt{4-r^2}}\right)}{\pi ^4 r^6 \left(r^2-4\right)^3 \left(r^2-2\right)^4} +\frac{\left(r^2-1\right)^3 \arctan^2\left(\frac{2 \sqrt{r^2-1}}{2-r^2}\right)}{\pi ^4 \left(r^2-2\right)^4}+\frac{\log ^2(r)}{\pi ^4} \,.
        \label{eq:self_interaction_Msq_sqrt_2_lt_r_lt_2}
    \end{multline}
    
    \item $r \geq 2$: 
    
    \begin{multline}
        \frac{\overline{\left| \mathcal{M} \right|^2}}{256 \pi^4 \alpha_D^4}  = 
        \log (r) \left[\frac{2 f(r) \log \left(\frac{2}{\sqrt{r^2-4}+r}\right)}{\pi ^4 r^3 \left(r^2-4\right)^{3/2} \left(r^2-2\right)^2}-\frac{4 \left(r^2-1\right)}{\pi ^4 r^2 \left(r^2-4\right)}-\frac{2 \left(r^2-1\right)^{3/2} \arctan\left(\frac{2 \sqrt{r^2-1}}{r^2-2}\right)}{\pi ^4 \left(r^2-2\right)^2}\right] \\
        + \sqrt{r^2-1} \left[\frac{4 \left(r^2-1\right)^2 \arctan\left(\frac{2 \sqrt{r^2-1}}{r^2-2}\right)}{\pi ^4 r^2 \left(r^2-4\right) \left(r^2-2\right)^2}-\frac{2 g(r) \log \left(\frac{2}{\sqrt{r^2-4}+r}\right) \arctan\left(\frac{2 \sqrt{r^2-1}}{r^2-2}\right)}{\pi ^4 r^3 \left(r^2-4\right)^{3/2} \left(r^2-2\right)^4}\right] \\
        -\frac{4 g(r) \log \left(\frac{2}{\sqrt{r^2-4}+r}\right)}{\pi ^4 r^5 \left(r^2-4\right)^{5/2} \left(r^2-2\right)^2} + \frac{(r^2 - 1)^3 \arctan^2 \left( \frac{2 \sqrt{r^2 - 1}}{r^2 - 2} \right)}{\pi^4 (r^2 - 2)^4} \\
        + \frac{4(r^2 - 1)^2}{\pi^4 r^4(r^2 - 4)^2} + \frac{f(r)^2 \log^2 \left(\frac{2}{\sqrt{r^2-4}+r}\right) }{\pi^4 r^6 (r^2 - 4)^3 (r^2 - 2)^4}+\frac{\log ^2(r)}{\pi ^4} \,.
        \label{eq:self_interaction_Msq_r_gt_2}
    \end{multline}
\end{enumerate}

where

\begin{alignat}{2}
    f(r) &\equiv&& \,\, r^{10} - 10r^8 + 26r^6 - 16r^4 + 16r^2 - 32 \,, \nonumber \\
    g(r) &\equiv&& \,\, r^{12} - 11r^{10} + 36r^8 - 42r^6 + 32r^4 - 48r^2 + 32 \,,\nonumber \\
    k(r) &\equiv&& \,\, r^{16} - 12r^{14} + 40r^{12} + 4r^{10} - 96r^8 - 224r^6 + 256r^4 + 256 \,,\nonumber \\
    p(r) &\equiv&& \,\, r^8 - 12r^6 + 36r^4 - 44r^2 + 20 \,,\nonumber  \\
    q(r) &\equiv&& \,\, r^{20} - 24r^{18} + 212r^{16} - 900 r^{14} + 2008 r^{12} \nonumber \\
    & &&\,\, - 2608 r^{10} + 2688 r^8 - 2432r^6 + 1280r^4 - 1024r^2 + 1024 \,.
\end{alignat}

\end{document}